\documentclass[a4paper,12pt]{article}
\usepackage{amssymb,amsmath,amsfonts}
\usepackage{verbatim}
\usepackage{graphicx}
\usepackage{epstopdf}
\newcommand{\Fig}[2]{%
\begin{center}
\parbox{10cm}{%
\refstepcounter{figure}\includegraphics[width=10cm]{#1}\\[12pt] \noindent Figure \thefigure:\quad
#2}\end{center}}
\newcommand{\TwoFig}[4]{%
\begin{flushleft}
\begin{tabular}{lr}
\parbox{7.7cm}{\includegraphics[width=7.7cm,height=6cm]{#1}}  & \parbox{7.7cm}{\includegraphics[width=7.7cm,height=6cm]{#3}} \\
\parbox{7.7cm}{\refstepcounter{figure}\noindent Figure \thefigure:\quad #2} & \parbox{8cm}{\refstepcounter{figure}\noindent Figure \thefigure:\quad #4}\\
\end{tabular}
\end{flushleft}
}
\newcommand{\TwoFigH}[4]{%
\begin{flushleft}
\begin{tabular}{lr}
\parbox{7.7cm}{\includegraphics[width=7.7cm,height=7.7cm]{#1}}  & \parbox{7.7cm}{\includegraphics[width=7.7cm,height=7.7cm]{#3}} \\
\parbox{7.7cm}{\refstepcounter{figure}\noindent Figure \thefigure:\quad #2} & \parbox{8cm}{\refstepcounter{figure}\noindent Figure \thefigure:\quad #4}\\
\end{tabular}
\end{flushleft}
}
\begin{document}
%%%%%%%%%%%%%%%%%%%%%%%%%
\begin{center}
{\bf \Large Qualitative Analysis and Numerical Simulation of Equations of the Standard Cosmological Model: $\Lambda\not=0$} \\[12pt]
Yurii Ignat'ev\\
N.I. Lobachevsky Institute of Mathematics and Mechanics, Kazan Federal University, \\ Kremleovskaya str., 35, Kazan, 420008, Russia
\end{center}

\begin{abstract}
On the basis of qualitative analysis of the system of differential equations of the standard cosmological model it is shown that in the case of zero cosmological constant this system has a stable center corresponding to zero values of potential and its derivative at infinity. Thus, the cosmological model based on single massive classical scalar field in infinite future would give a flat Universe. The carried out numerical simulation of the dynamic system corresponding to the system of Einstein - Klein - Gordon equations showed that at great times of the evolution the invariant cosmological acceleration has an oscillating character and changes from $-2$ (braking), to $+1$ (acceleration). Average value of the cosmological acceleration is negative and is equal to $-1/2$. Oscillations of the cosmological acceleration happen on the background of rapidly falling Hubble constant. In the case of nonzero value of the cosmological constant depending on its value there are possible three various qualitative behavior types of the dynamic system on 2-dimensional plane  $(\Phi,\dot{\Phi})$, which correspond either to zero attractive focus or to stable attractive knot with zero values of the potential and its derivative. Herewith the system asymptotically enters the secondary inflation.
Carried out numerical simulation showed that at cosmological constant $\Lambda<m^2 3\cdot10^{-8}$ the macroscopic value of the cosmological acceleration behaves itself similar to the case $\Lambda=0$, i.e. in the course of the cosmological evolution there appears a lasting stage when this value is close to $-1/2$ which corresponds to non-relativistic equation of state. In this article, the results of qualitative and numerical analysis, obtained in \cite{Yu_Quality}, common to the case of a non-zero cosmological term.\\[12pt]
{\bf keyword} standard cosmological model, instability, quality analysis, numerical simulation, zero - center flat universe, cosmological term, numerical gravitation.
\end{abstract}

This work was funded by the subsidy allocated to Kazan Federal University for the state assignment in the sphere of scientific activities.

\section{Basic Relations of the Standard Cosmological Model}
Let us consider the space-flat Friedman Universe
\begin{equation}\label{Friedmann}
ds^2=dt^2-a^2(t)d\ell^2_0
\end{equation}
($d\ell^2_0$ is a metrics of 3-dimensional Euclidean space), generated by homogenous massive scalar field whose potential $\Phi(t)$ complies with the Klein - Gordon equation:
\begin{equation}
\label{d2F(t)}
\ddot{\Phi}+3\frac{\dot{a}}{a}\dot{\Phi}+m^2\Phi=0
\end{equation}
where the derivative of $t$ is denoted with a dot. Herewith the scale factor $a(t)$ complies with a unique non-trivial Einstein equation:
\begin{equation}\label{Einst_Eq_S}
3\frac{\dot{a}^2}{a^2}=\Lambda+8\pi(\dot{\Phi}^2+m^2\Phi^2),
\end{equation}
where $\Lambda$ is a cosmological term which we further suppose non-negative:
\begin{equation}\label{Lambda>0}
\Lambda\geq0.
\end{equation}
These well known equations\footnote{see e.g. \cite{Gorb_Rubak}.} (\ref{d2F(t)}) and (\ref{Einst_Eq_S}) represent a system of autonomous ordinary differential equations the analysis of which is a subject of this research.

Let us introduce the further required scalar functions:
Hubble constant
\begin{equation}\label{H}
H(t)=\frac{\dot{a}}{a}
\end{equation}
and invariant cosmological acceleration:
\begin{equation}\label{Omega}
\Omega(t)=\frac{\ddot{a}a}{\dot{a}^2}\equiv 1+\frac{\dot{H}}{H^2}
\end{equation}
Let us note the following useful relation between $\Omega$ and barotropic coefficient $\kappa$ in the equation of state $p=\kappa\varepsilon$:
\begin{equation}\label{Omega-kappa}
\Omega = -\frac{1}{2} (1+3\kappa).
\end{equation}
Let us also write down the expressions for the scalar field's energy density and pressure:
\begin{equation}
\label{eps}
\varepsilon=\dot{\Phi}^2+m^2\Phi^2; \quad p=\dot{\Phi}^2-m^2\Phi^2,
\end{equation}
so that
\begin{equation}\label{e+p}
\varepsilon+p=2\dot{\Phi}^2,
\end{equation}
\section{Qualitative Analysis of the Dynamic System of the Standard Cosmological Model}
\subsection{Reduction of the System of Equations to the Canonical Form}
First, let us carry out scaling of the equations proceeding to new dimensionless time variable $\tau$
\begin{equation}\label{tau}
\tau=mt,\Rightarrow \dot{f}=mf',
\end{equation}
where $f'=df/d\tau$. Thus, from (\ref{d2F(t)}) and (\ref{Einst_Eq_S}) we get:
\begin{eqnarray}
\label{Phi''}
\Phi''+3H_m\Phi'+\Phi=0;\\
\label{Einst_h}
3H^2_m=\Lambda_m+8\pi \bigl(\Phi'^2+\Phi^2\bigr),
\end{eqnarray}
where $H_m(\tau)$ and $\Lambda_m$ are the Hubble constant and cosmological constant, both measured in Compton time units:
\begin{equation}\label{h}
H_m(\tau)=\frac{a'}{a}=\frac{H}{m}; \quad \Lambda_m=\frac{\Lambda}{m^2}.
\end{equation}
Let us notice that equations (\ref{Phi''}) and (\ref{Einst_h}) represent a system of ordinary nonlinear differential equations. Taking into account condition
\begin{equation}\label{dota>0}
\dot{a}\geq 0 \Leftrightarrow H\geq 0
\end{equation}
and using a standard substitution this system can be reduced to the form of \emph{normal autonomous system of ordinary differential equations on a plane}:
\begin{eqnarray}\label{Z}
\Phi'&=&Z(t);\\
\label{Z'}
Z'&=&-3H_m Z-\Phi,
\end{eqnarray}
where function $H_m(\Phi,Z)$ is \emph{algebraically} defined through the Einstein equation using functions $\Phi(\tau)$ and $Z(\tau)$:
\begin{equation}\label{h(tau)}
H_m=\frac{1}{\sqrt{3}}\sqrt{\Lambda_m+8\pi\bigl(Z^2+\Phi^2\bigr)}.
\end{equation}
Thus, we finally get a system of autonomous differen\-tial equations of the dynamic system on the plane $(\Phi,Z)$:
\begin{equation}\label{eqs0}
\left\{\begin{array}{l}
\Phi'=Z(t);\\
Z'=-\sqrt{3\pi}\sqrt{\Lambda_m+8\pi\bigl(Z^2+\Phi^2\bigr)}Z-\Phi;
\\
\end{array}\right.
\end{equation}
or, in terms of qualitative theory of ordinary differen\-tial equations (see e.g. \cite{Bogoyavlensky}):
\begin{equation}\label{eqs}
\left\{\begin{array}{l}
\displaystyle\frac{dx}{dt}=P(x,y); \\[12pt]
\displaystyle\frac{dy}{dt}=Q(x,y)\\[12pt]
\end{array}\right.
\end{equation}
where
\begin{eqnarray}\label{P,Q}
x\equiv \Phi;\;y\equiv Z;\; P(x,y)\equiv y;\; \nonumber\\
Q(x,y)\equiv -\sqrt{3\pi}\sqrt{\Lambda_m+8\pi\bigl(x^2+y^2\bigr)}y-x.
\end{eqnarray}
This system of equations can be investigated and asymptotic behavior of the solutions at $t\to\pm\infty$ can be defined with a help of qualitative theory of differential equations. The next property of the Standard Cosmological Model (SCM) is an important one:

\emph{The evolution of the Universe in the SCM with cosmological term in terms of time variable $\tau$ is defined by the only one parameter $\Lambda_m$ and initial conditions.}

\subsection{Singular Points of the Dynamic System}
Singular points of the dynamic system $M_0(x_0,y_0$ (\ref{eqs}) are defined by zeroes of the derivatives (see e.g. \cite{Bogoyavlensky}):
\[P(x_0,y_0)=0;\quad Q(x_0,y_0)=0.\]

It is not difficult to see that the dynamic system (\ref{eqs}) as in the case $\Lambda\equiv0$ has a unique singular point:

\begin{equation}\label{Phi0,Z0}
M_0=(0,0)\longleftrightarrow x_0=\Phi_0=0;\; y_0=Z_0=0.
\end{equation}
\subsection{A Kind of Singular Point}
To define a kind of a singular point it is necessary to find the eigenvalues of the characteristic polynomial:
\begin{equation}\label{haracter_pol}
\Delta(\lambda)=\left|
\begin{array}{ll}
P'_x(x_0,y_0)-\lambda & P'_y(x_0,y_0)\\[12pt]
Q'_x(x_0,y_0) & Q'_y(x_0,y_0)-\lambda\\
\end{array}
\right|=0,
\end{equation}
where partial derivatives of functions $P(x,y)$, $Q(x,y)$ are calculated in a singular point $M_0$. Calculating derivatives of functions $P,Q$ in (\ref{P,Q}), let us find:
\[P'_x(0,0)=0;\quad Q'_y(0,0)=1;\]
\[Q'_x(0,0)=-1;\quad Q'_y(0,0)=-\sqrt{3\Lambda_m}\]
Thus, the characteristic polynomial (\ref{haracter_pol}) is equal to:
\[
\Delta(\lambda)=\left|
\begin{array}{rr}
-\lambda & 1\\[12pt]
-1 & -\lambda-\sqrt{3\Lambda_m}\\
\end{array}
\right|=0,
\]
where from we find its roots
\begin{equation}\label{lambda}
\lambda_\pm=-\frac{1}{2}\sqrt{3\Lambda_m}\pm\frac{1}{2}\sqrt{3\Lambda_m-4}.
\end{equation}
Eigenvalues satisfy the following identity:
\begin{equation}\label{l1l2}
\lambda_1\lambda_2\equiv1.
\end{equation}
Thus, four essentially different cases are possible (see \cite{Bogoyavlensky}):

\noindent
1. The case of zero cosmological term:
\begin{equation}\label{Lambda<4/3}
\Lambda_m\equiv0
\end{equation}
we have two complex conjugated imaginary eigen\-values:
\begin{equation}\label{Re=0}
\lambda_\pm=\pm i.
\end{equation}
Since eigenvalues turned to be pure imaginary ones then \emph{a unique singular point} (\ref{Phi0,Z0}) of the dynamic system (\ref{eqs}) is its center (see \cite{Bogoyavlensky}). In this case at $\tau\to+\infty$ the phase trajectory of the dynamic system is winded round this center making an infinite number of turns.\\

\noindent
2. The case of small cosmological term:
\begin{equation}\label{Lambda<4/3}
0<\Lambda_m<\frac{4}{3}
\end{equation}
we have two complex conjugated eigenvalues and it is
\begin{equation}\label{Re<0}
Re(\lambda)=-\frac{\sqrt{3\Lambda_m}}{2}<0.
\end{equation}
In this case in accordance with qualitative theory of differential equations, point $M_0$ (\ref{Phi0,Z0}) is \emph{an attractive focus}, all phase trajectories of the dynamic system at $\tau\to+\infty$ are twisting spirals, which are winded round the singular point performing an infinite number of turns. This case effectively coincide quali\-ta\-ti\-vely with the previous one.\\

\noindent
3. The case of great value of the cosmological term:
\begin{equation}\label{Lambda>4/3}
\Lambda_m>\frac{4}{3}
\end{equation}
-- then we have two various real and, according to (\ref{lambda}), negative eigenvalues $\lambda_1\not=\lambda_2$, $\lambda_1<0,\lambda_2<0$. In such a case the singular point is \emph{a stable attractive knot}. At $\tau\to+\infty$ all phase trajectories of the dynamic system enter the singular point and all all the trajectories apart two exceptional ones, when coming to this singular point, are tangent to eigenvector $\mathbf{u}_1$, which corresponds to eigenvalue, being minimal by its module, i.e. $\lambda_1$. Two exceptional trajectories are tangent to the second eigenvector $\mathbf{u}_2$. Mentioned eigenvectors are equal to:
\begin{equation}\label{u1u2}
\mathbf{u}_1=(1,\lambda_1);\quad \mathbf{u}_1=(1,\lambda_2).
\end{equation}
The angle $\alpha$ between the eigenvectors is defined by means of the relation:
\begin{equation}\label{alpha}
\cos\alpha\equiv \frac{\mathbf{u}_1\mathbf{u}_2}{\sqrt{\mathbf{u}^2_1\mathbf{u}^2_2}}=\sqrt{\frac{4}{3\Lambda_m}}<1.
\end{equation}
At very large values of $\Lambda_m$ the angle between the eigenvectors tends to $\pi/2$, at $\Lambda_m\to4/3$ this angle tends to zero.\\

\noindent
4. The degeneration case:
\begin{equation}\label{Lambda=4/3}
\Lambda_m=\frac{4}{3}
\end{equation}
-- this case effectively coincides with the previous one with an account of the circumstance that all trajectories enter the singular point tangent to unique eigenvector -- this exactly corresponds to mentioned above extreme case $\alpha\to0$.
Thus, the phase trajectory of the dynamic system based on the equation of classical massive scalar field (\ref{d2F(t)}) and the Einstein equation (\ref{Einst_Eq_S}), in the plane $(\Phi,Z)$ has a single zero singular point (attractive focus or attractive stable point) (\ref{Phi0,Z0}), where it is:
\begin{equation}\label{center}
t\to+\infty \Rightarrow \Phi\to0;\quad \dot{\Phi}\to 0\Rightarrow H\to \sqrt{\frac{\Lambda}{3}},
\end{equation}
\begin{equation}
\Omega\to\left\{\begin{array}{ll}
 1, & \Lambda\not\equiv 0;\\
 0, & \Lambda \equiv 0.
\end{array}
\right.
\end{equation}
All that is changing is a type of the singular point and along with that the details of the approach of the phase trajectories to the singular point $\Phi=0$, $\dot{\Phi}=0$ at $\tau\to+\infty$.

\subsection{The Asymptotic Behavior of the Scale Factor}
Since at $\tau\to+\infty$ (or $t\to +\infty$) $\Phi\to0$ and $\dot{\Phi}\to 0$, then in the absence of other forms of matter apart from the scalar field, the Universe stays alone with the $\Lambda$ - term, therefore in consequence of the Einstein equation (\ref{Einst_Eq_S}) the scale factor evolves by the inflation law:
\begin{equation}\label{inflat2}
a(t)\sim \mathrm{e}^{H_0 t},\quad t\to+\infty,
\end{equation}
where $H_0$ is the Hubble constant:
\begin{equation}\label{H_0}
H_0=\sqrt{\frac{\Lambda}{3}}\equiv m\sqrt{\frac{\Lambda_m}{3}}.
\end{equation}
At early stages it is $\tau\to-\infty$, until:
\begin{equation}\label{t_inflat1}
\Phi(\tau)\approx \Phi_0=\mathrm{Const}; \longrightarrow 8\pi\Phi^2_0\gg \Lambda_m,
\end{equation}
\begin{equation}\label{H_1}
H(t)\approx H_1=\frac{m}{\sqrt{3}}\sqrt{\Lambda_m+8\pi\Phi^2_+0}>H_0
\end{equation}
early inflation takes place:
\begin{equation}\label{inflat1}
a(t)\sim \mathrm{e}^{H_1 t},\quad t\to-\infty
\end{equation}

Thus, the invariant cosmological acceleration
\begin{equation}\label{Omega}
\Omega=\frac{\ddot{a}a}{\dot{a}^2}\equiv 1+\frac{\dot{H}}{H^2}
\end{equation}
at $\Lambda\not\equiv0$ tends to one at early and late stages:
\begin{equation}\label{Omega+-8}
\Omega(t)\to 1, \quad t\to\pm\infty.
\end{equation}
Along with this, the Hubble ``constant'' has constant values in these extreme cases:
\begin{eqnarray}\label{H1H0}
H(t)\to H_1;\; (t\to -\infty),\nonumber\\
H(t)\to H_0;\; (t\to+\infty); \quad (H_1>H_0).
\end{eqnarray}

\subsection{The Phase Trajectories of the Dynamic System (\ref{eqs0})}
It is necessary to bear in mind that time variable on all the plots is $\tau$, i.e. the time measured at Compton scale. Since the evolution of the investigated dynamic system (\ref{eqs}) is defined only by initial conditions and the value of normalized cosmological constant $\Lambda_m$, let us consider the dependency of the evolution details on the value of the cosmological constant and the initial conditions. Further, let us accept the assumption that $\dot{\Phi}(-\infty)=0$. We will investigate the properties of phase trajectories in terms of the plot shown on Fig. \ref{ris1}:
\Fig{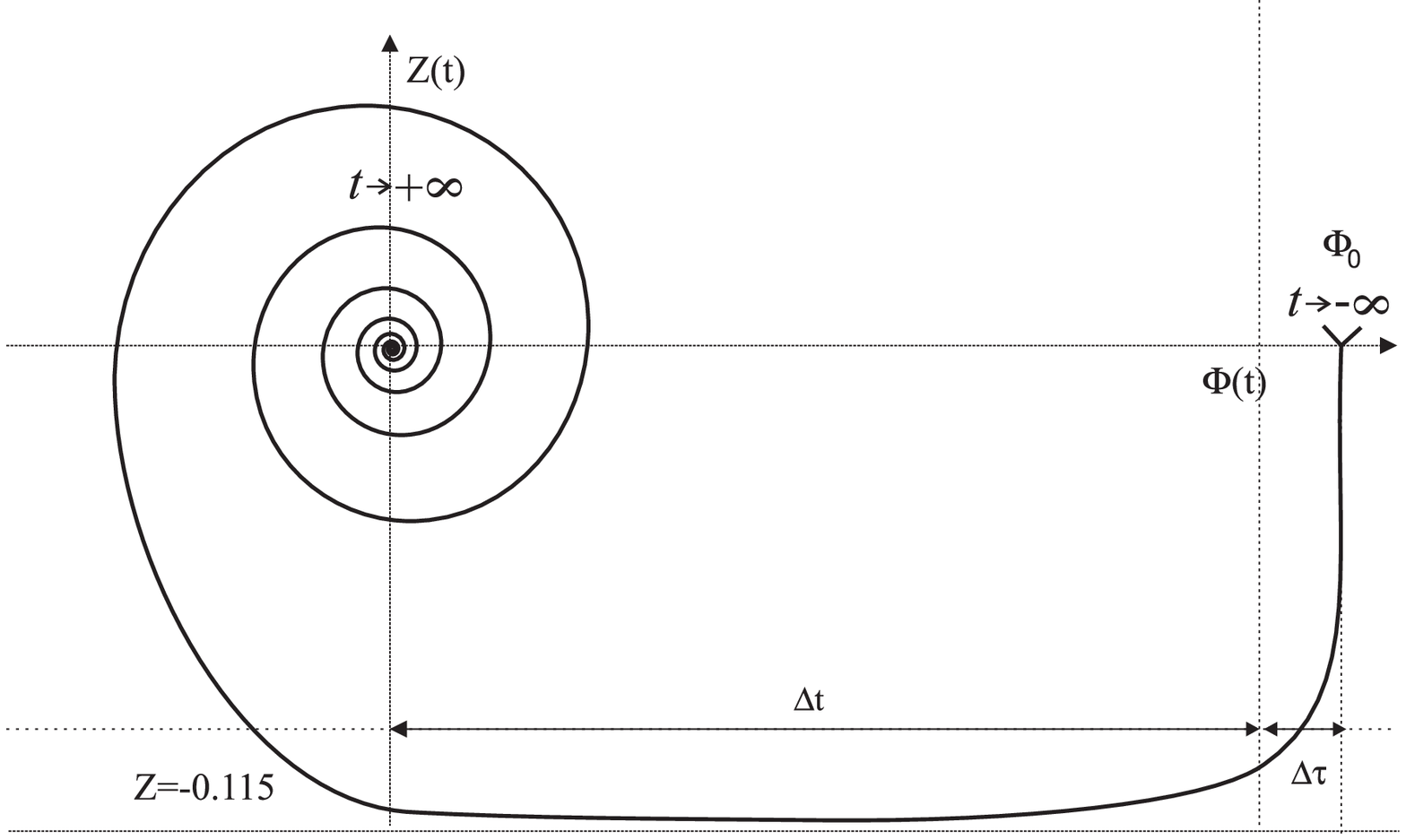}{\label{ris1}The qualitative view of the phase trajectory of the dynamic system (\ref{eqs0}) at $\Lambda_m\ll1$. On this figure $\Delta\tau$ is a characteristic time of decrease of the potential rate of change till the ``bottom'' of  the plot, $Z_0\approx-0.115$, $\Delta t$ is a characteristic time of decrease of the potential value with a constant velocity $\Phi'\approx Z_0$. After this instant of time there start the winding of the phase trajectory round the zero center. The number of spiral turns is infinite.}
\noindent
1. The initial stage with a duration $\Delta\tau$ with $\Phi\approx\Phi_0$ -- right part of the plot; this stage is characterized by a rapid fall of the potential's derivative from $0$ to  ``enigmatic number'' $-0.115$. Actually there is no any enigma in this number (see (\ref{eqs0})):
\begin{equation}\label{Z0}
Z_0=-\frac{1}{\sqrt{24\pi}}\approx - 0.1151647165.
\end{equation}
Actually, inflation happens at this stage.\\
\noindent

2. The Middle stage with duration $\Delta t$ is a medium part of the plot; $Z=\Phi'\approx \mathrm{Const}=Z_0$ at this stage. Potential falls to significantly small values at this stage.\\

\subsection{Zero Value of $\Lambda$: $\Lambda_m=0$}
\TwoFig{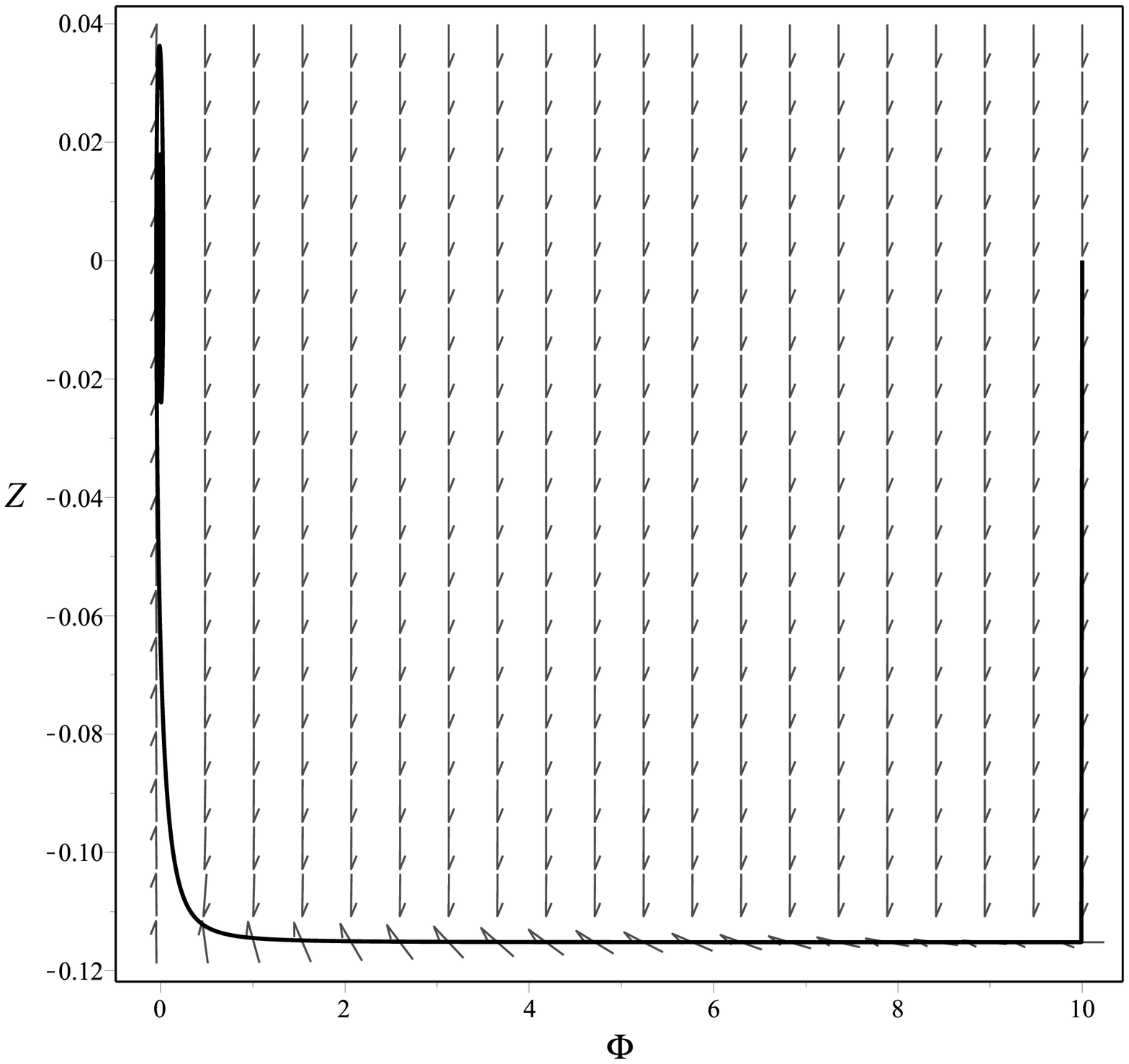}{\label{ris2}The large-scale phase portrait of the dynamic system (\ref{eqs}) $\tau\in[-1000,1000]$; $\Phi(-1000)=10$, $Z(-10000)=0$.}{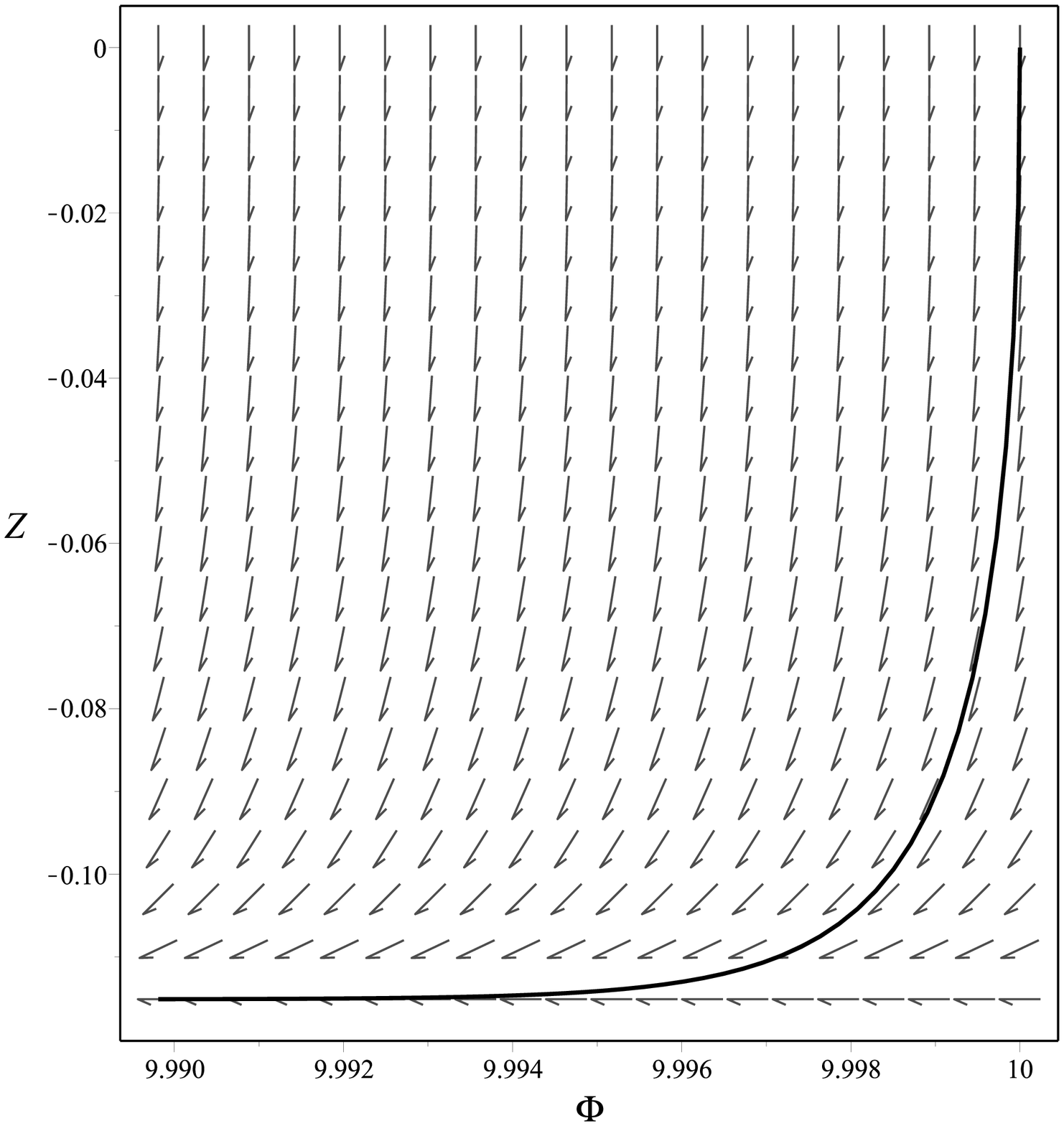}{\label{ris3}The initial stage of descent of the dynamic system (\ref{eqs0}) (the right-most part of the plot on Fig. \ref{ris1}) $\tau\in[-1000,-999.9]$; $\Delta\tau\lesssim 10^{-1}$; $\Phi(-1000)=10$, $Z(-10000)=0$.}
\noindent
3. The final stage of evolution with infinite duration; damped oscillations of the potential and its deri\-va\-tive happen at this stage.
The Universe stays asympto\-ti\-cally flat.

\TwoFig{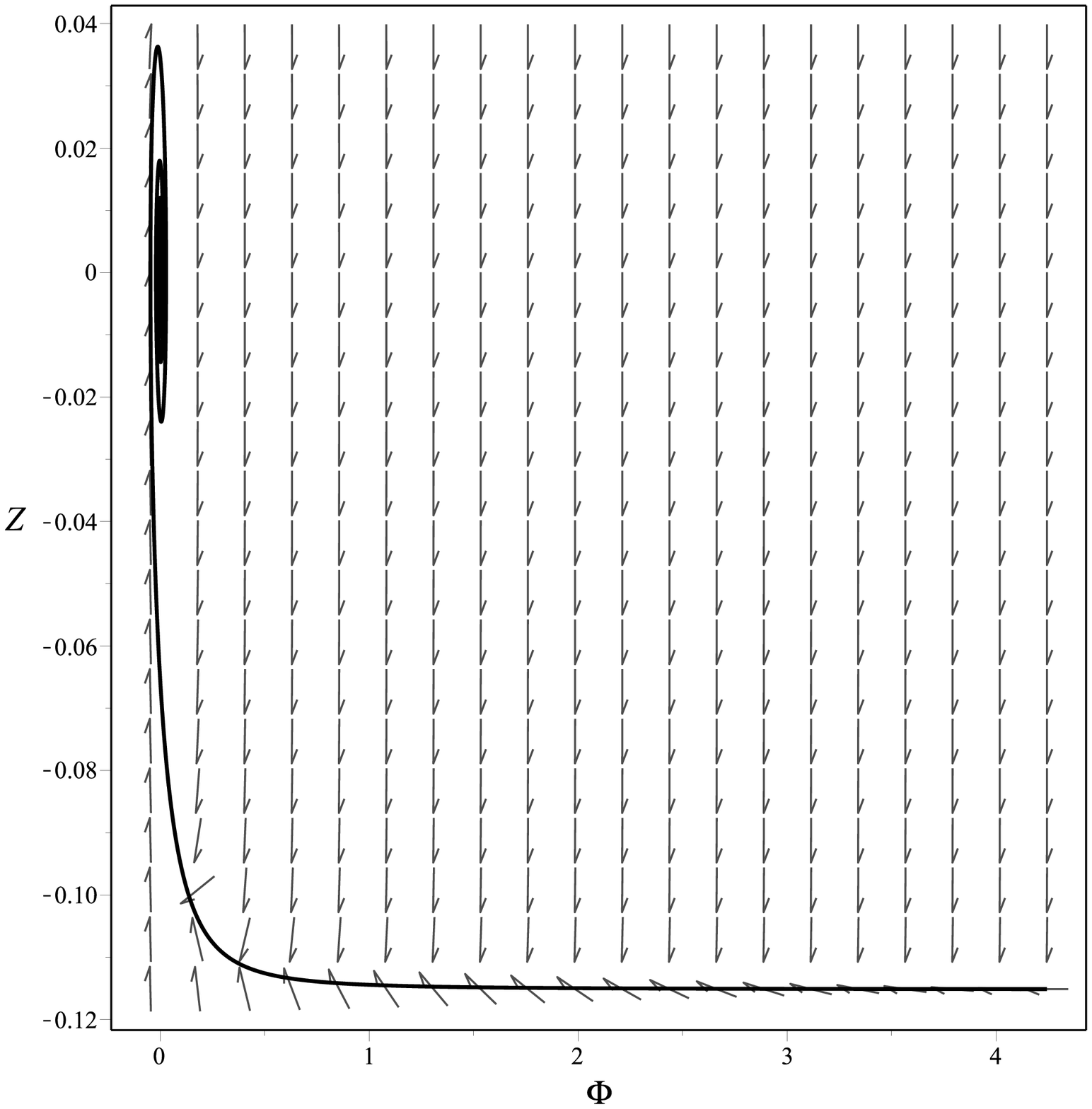}{\label{ris4}The middle stage of the dynamic system (\ref{eqs0}) $\Phi'\approx \mathrm{Const}\approx -0.115$ $\tau\in[-950,100]$;  $\Phi(-1000)=10$, $Z(-10000)=0$.}{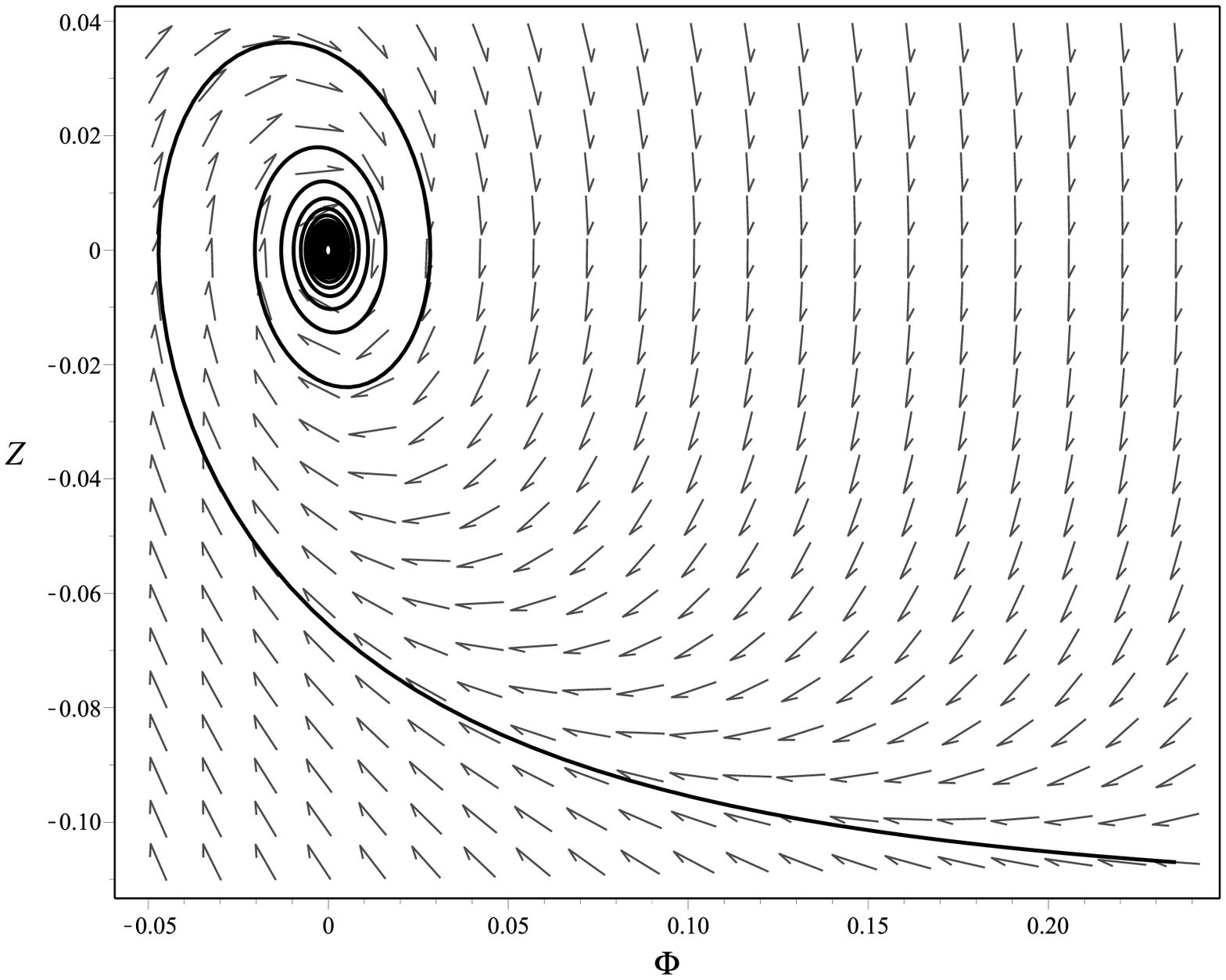}{\label{ris5}Winding round the center $M_0=(0,0)$ of the dynamic system (\ref{eqs0}) (the left-most part of the plot shown on Fig. \ref{ris1}) $\tau\in[-915,-700]$; $\Phi(-1000)=10$, $Z(-10000)=0$. }

The plots below show the results of numerical simulation of the dynamic system (\ref{eqs0}) at various initial conditions. In this paper it is used a Rosenbrock's method well adapted to integration of stiff systems of differential equations.
Let us show the characteristic examples of phase portraits of the dynamic system (\ref{eqs0}), obtained using numerical simulation methods in the system of the applied mathematics Maple XVII.

\TwoFigH{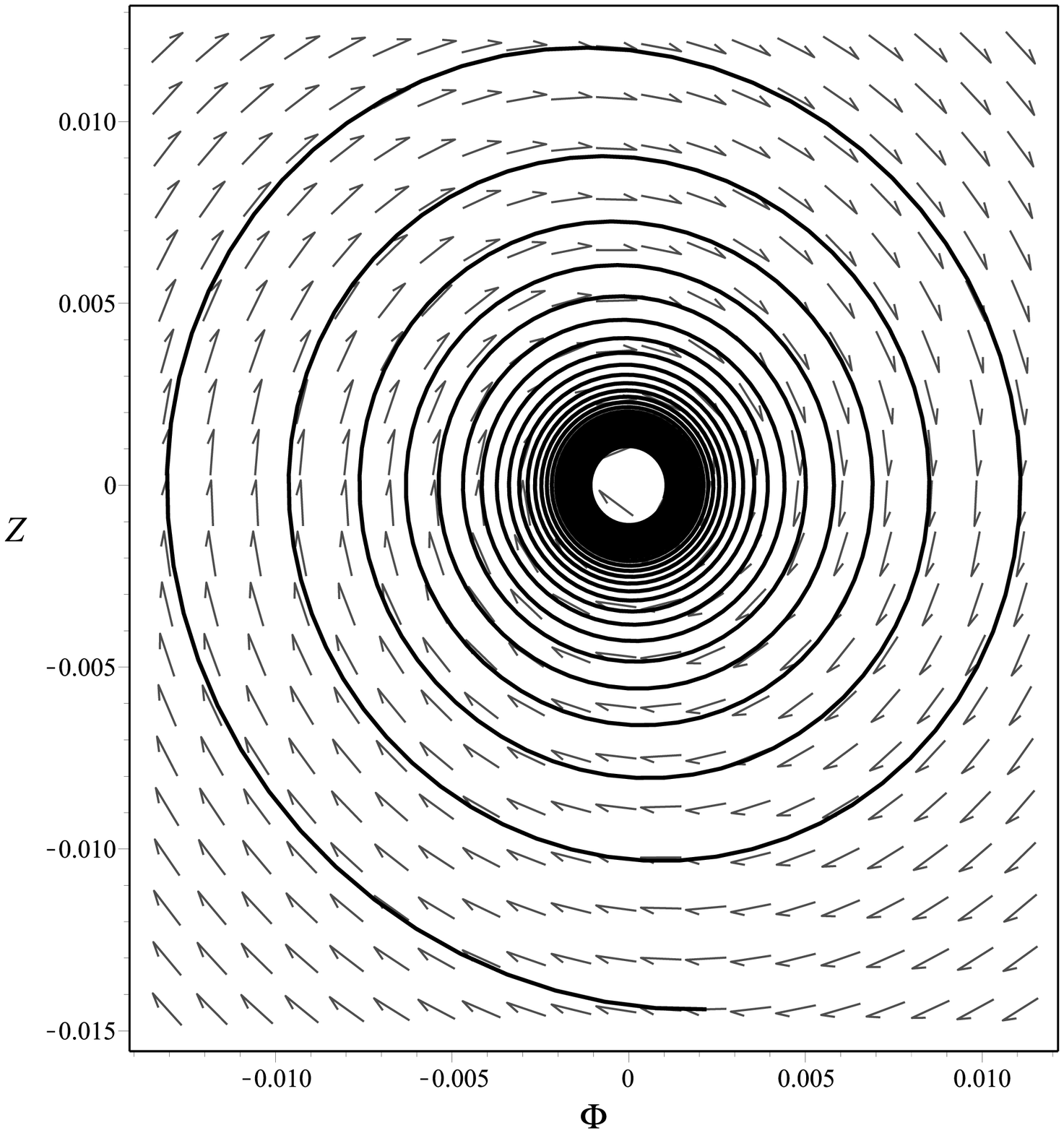}{\label{ris6}The final stage of the dynamic system (\ref{eqs0}): winding round the center $M_0=(0,0)$ at initial conditions: $\Phi(-1000)=10,\dot{\Phi}(-1000)=0$; $\tau\in[-900,-700]$.}{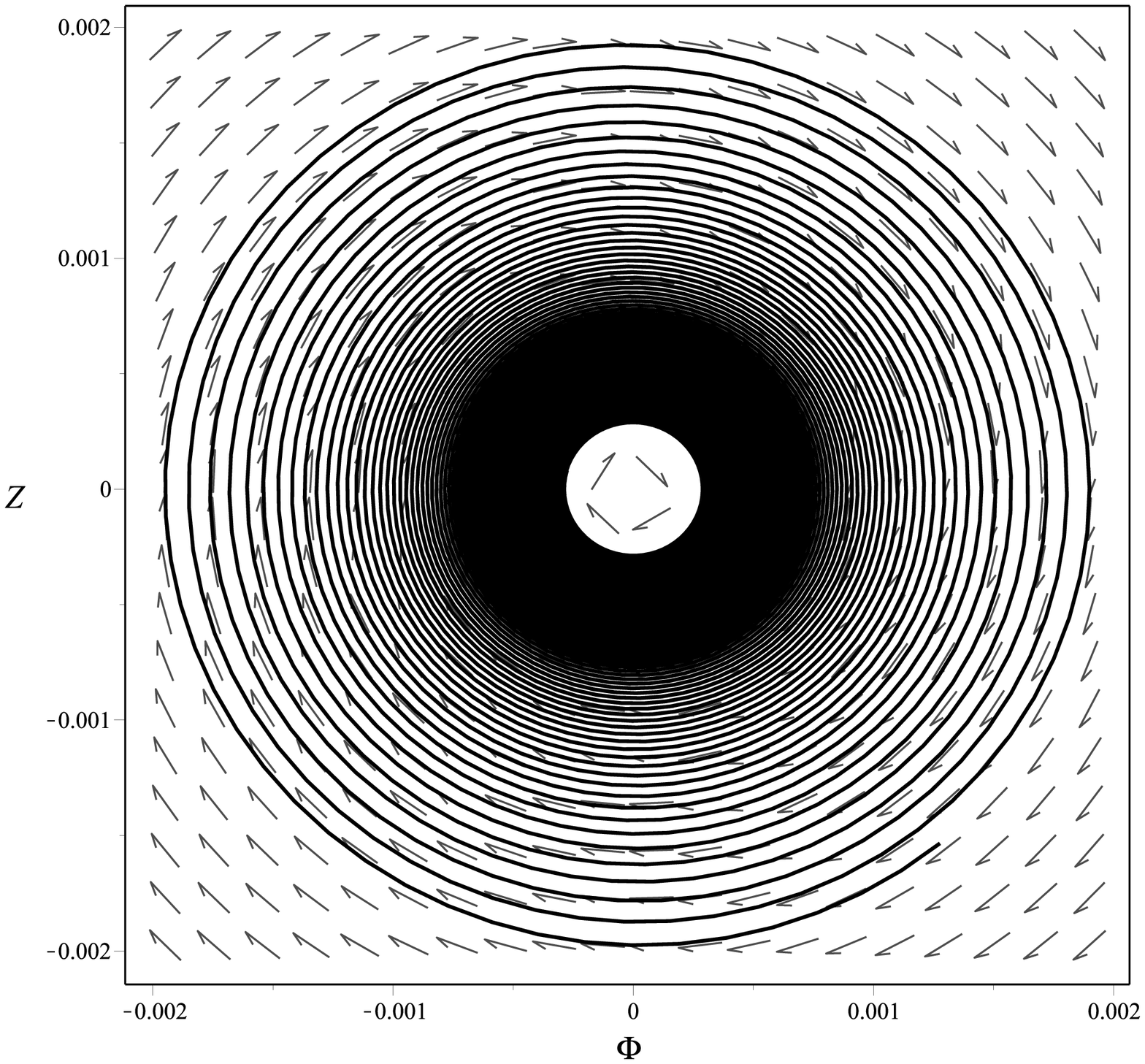}{\label{ris7}The final stage of the dynamic system (\ref{eqs0}): winding round the center $M_0=(0,0)$ of the dynamic system (\ref{eqs0}) (left part of the plot shown on Fig. \ref{ris1}) $\tau\in[-800,-100]$. }

Since characteristic properties of the system's (\ref{eqs0}) phase portraits have incomparable scales, we show fragments of phase planes on different time intervals.

\subsubsection{Small Values of $\Lambda$: $\Lambda_m=0.1<\frac{4}{3}$}
In this case the final point of the phase trajectory is the attractive focus (\ref{Phi0,Z0}) $\Phi_0=0;\; y_0=Z_0=0.$
\TwoFig{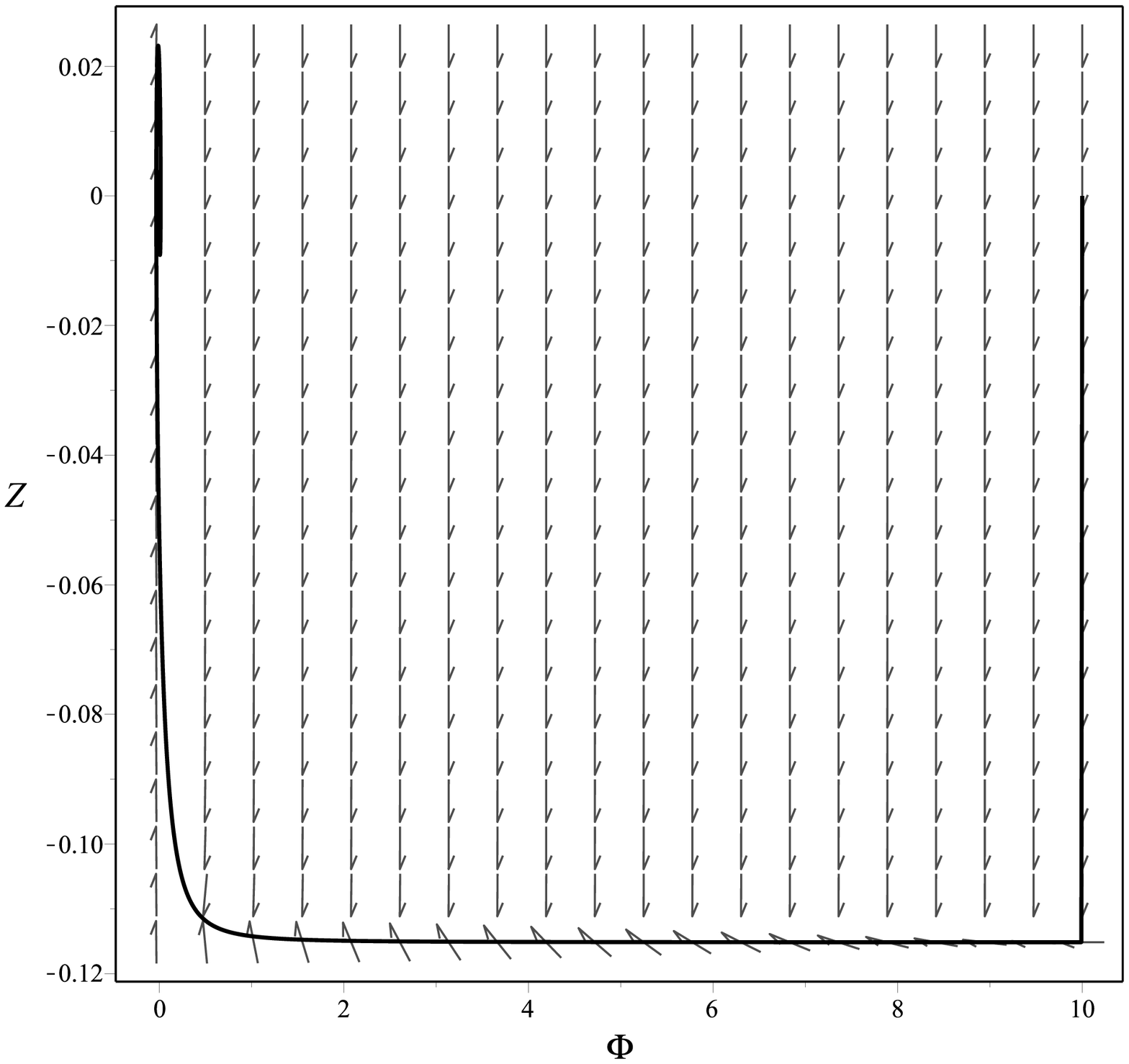}{\label{ris8}The large-scale picture of the phase trajectory $\tau\in[-1000,1000]$ at initial values: $\Phi(-1000)=10$; $Z(-1000)=0$.}{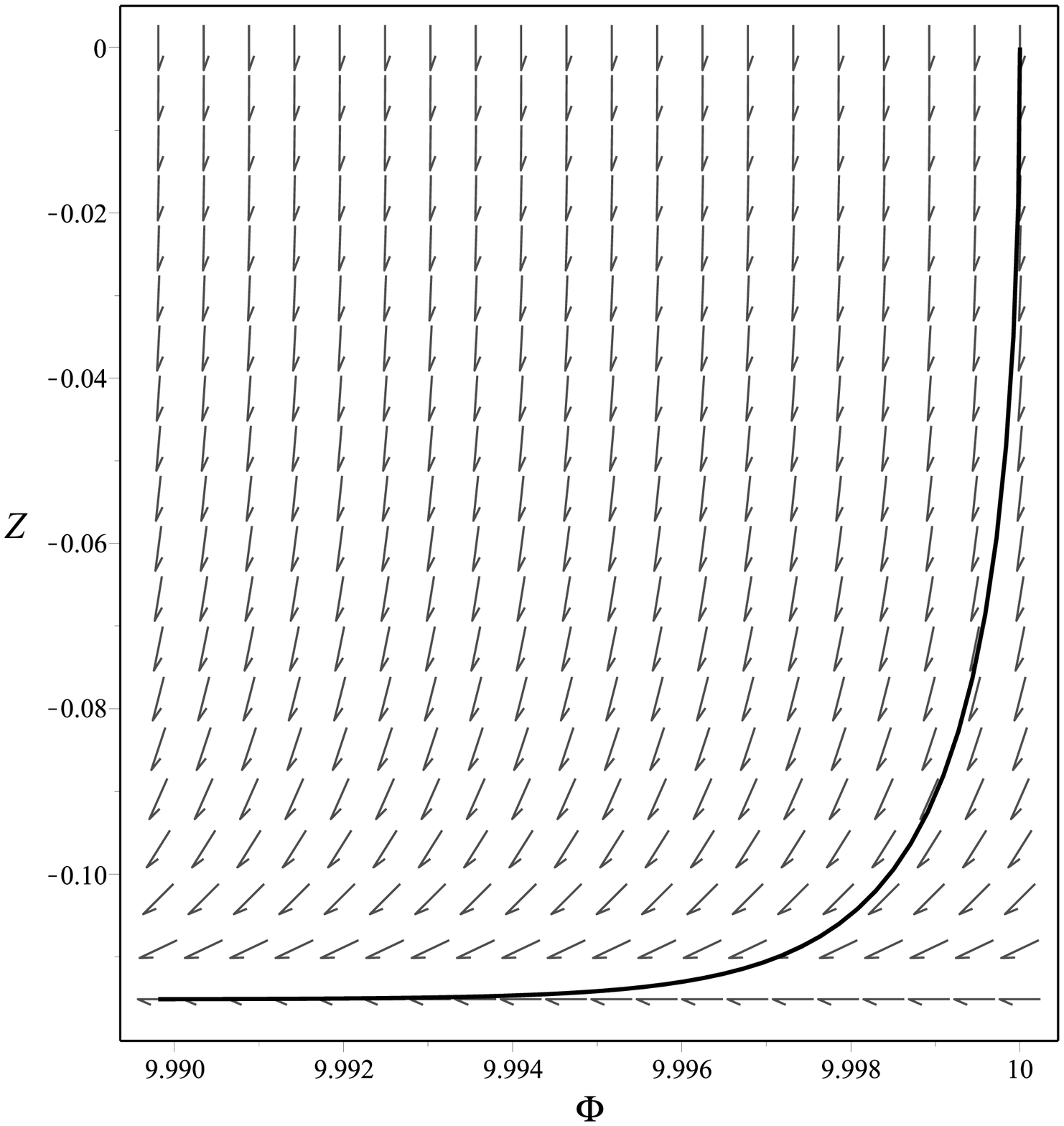}{\label{ris9}The stage of the phase trajectory's descent at times $\tau\in[-1000,-999.9]$ at initial values: $\Phi(-1000)=10$; $Z(-1000)=0$. }
\TwoFig{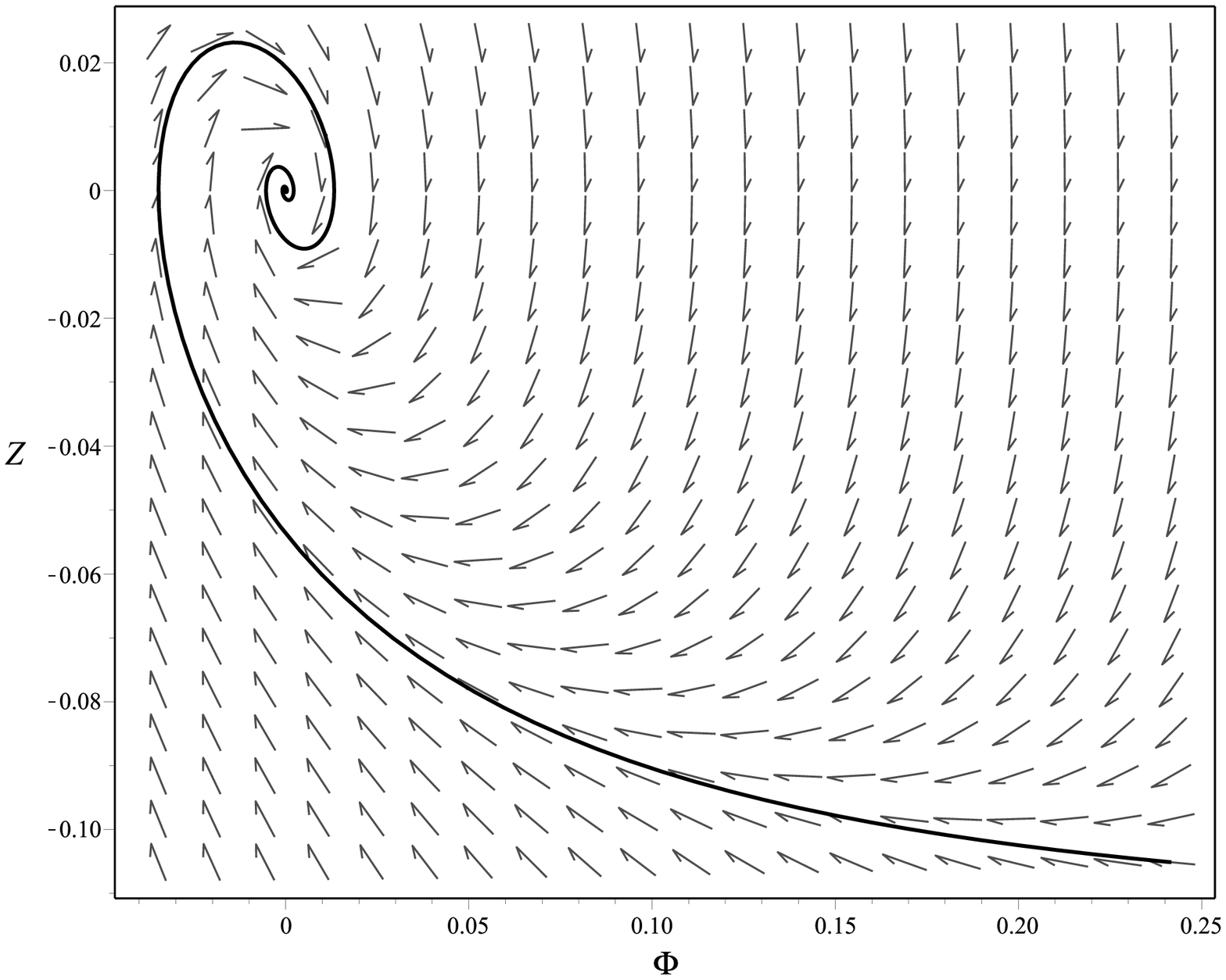}{\label{ris10}The stage of winding of the phase trajectory round the attractive focus: $\tau\in[-915,-700]$ at initial values: $\Phi(-1000)=10$; $Z(-1000)=0$. }{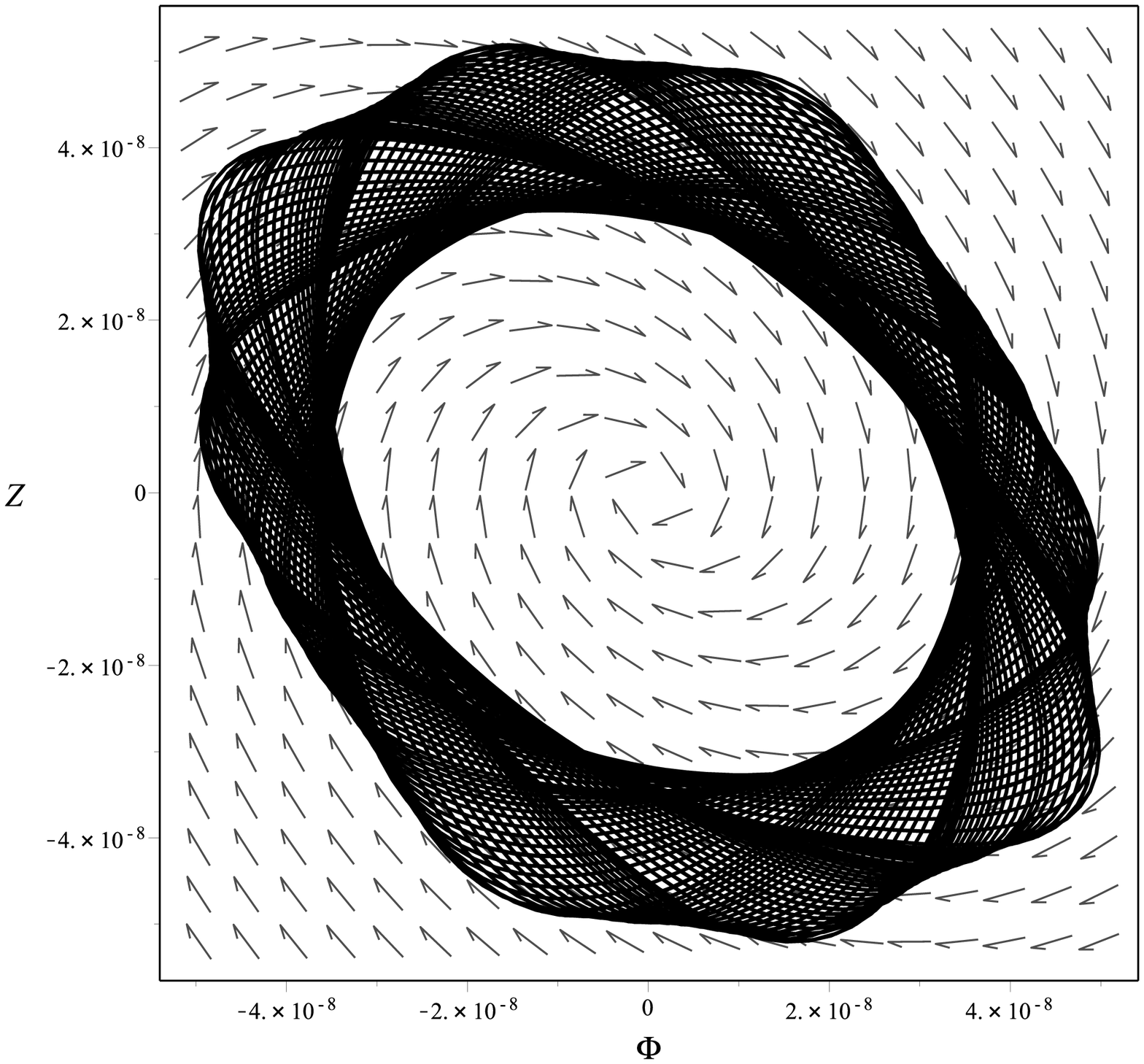}{\label{ris11}The final stage of the phase trajectory -- twisting spiral line: $\tau\in[-800,-100]$ at initial values: $\Phi(-1000)=10$; $Z(-1000)=0$.}

\subsubsection{Large Values of $\Lambda$: $\Lambda_m=10>\frac{4}{3}$}
In this case the final point of the phase trajectory is the stable attractive knot (\ref{Phi0,Z0}) $\Phi_0=0;\; y_0=Z_0=0.$
The large-scale picture of the phase trajectory does not qualitatively differ from the previous one however the method of entry to this point is changed (Fig. \ref{ris12} -- \ref{ris13}).
\TwoFig{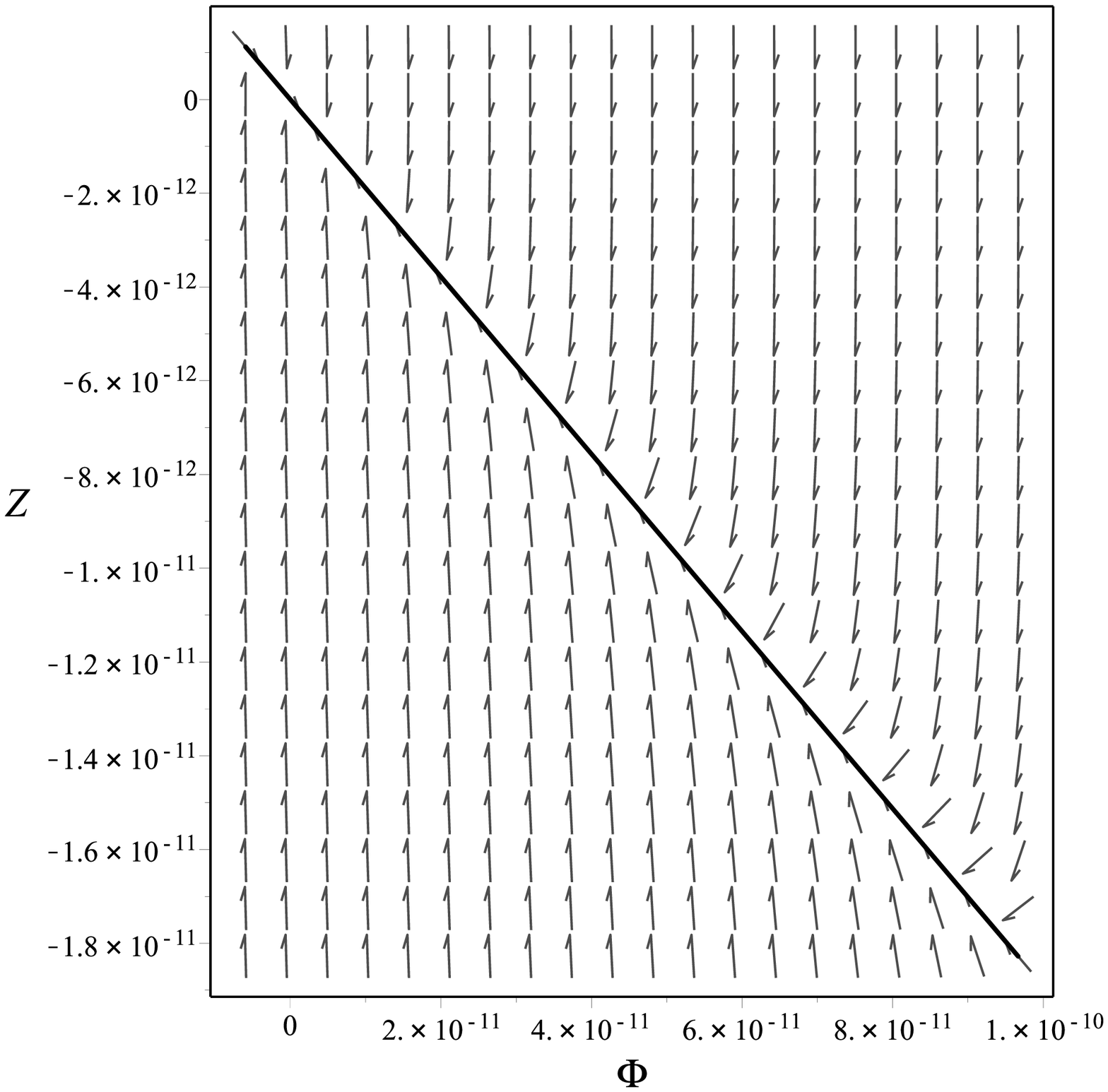}{\label{ris12}Approach to the attractive knot of the phase trajectory: $\tau\in[85,1000]$ at initial values: $\Phi(-1000)=10$; $Z(-1000)=0$.}{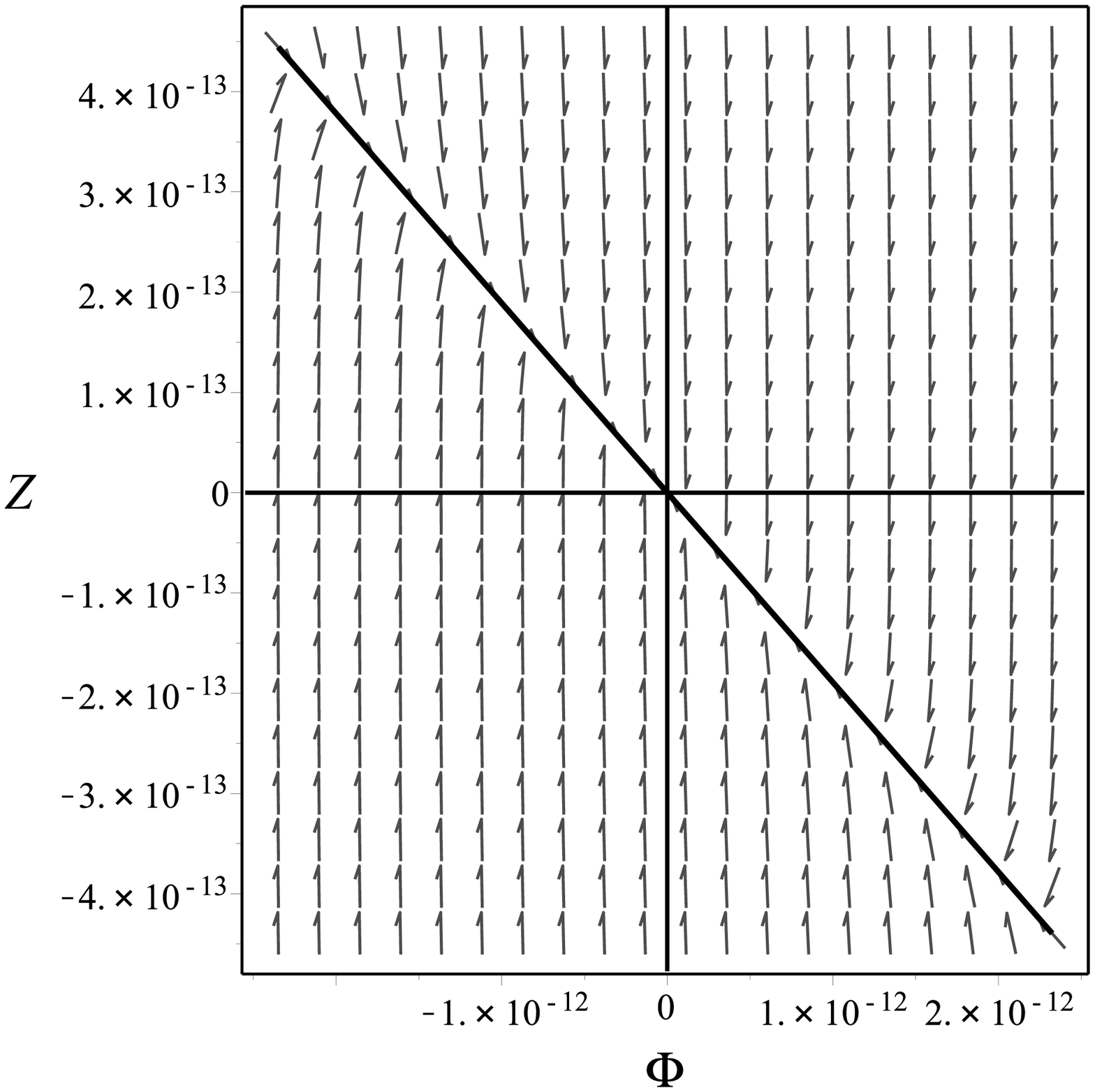}{\label{ris13}Passing through the attractive knot of the phase trajectory: $\tau\in[400,600]$ at initial values: $\Phi(-1000)=10$; $Z(-1000)=0$.}

\section{Numerical Integration of the Dynamic Equations}
However, phase portraits of the dynamic system (\ref{eqs0}) presented on Fig. \ref{ris2} -- \ref{ris13} do not provide infor\-ma\-tion about certain details of the cosmological evolution which are possible to obtain only by direct numerical integration of the original system of Einstein - Klein - Gordon equations.
Moreover, they do not provide information about original 3-dimensional dynamic system which also includes the Einstein equation (\ref{Einst_Eq_S}) as well as they do not provide information about observed cosmological scalars $H(t)$ and $\Omega(t)$. We should apply methods of direct numerical integrations of the original system of differential equations to obtain this information. Below we show the results of numerical integration of these equations using Rosenbrock's method, which is well adapted for numerical  integration of systems of ordinary diffe\-ren\-tial equations possessing the stiffness criteria. In series of more simple cases we used the standard Runge - Kutta - Fehlberg method of 4-5 orders and in series of more complex cases- the Runge-Kutta method of 7-8 orders.
\subsection{The Evolution of the Potential and its Derivative}
At the initial stages $\tau\to\-\infty$ at large enough values of the potential the scalar field evolves the same way as in the case $\Lambda=0$: the value of the scalar field's potential falls linearly with time. Then the transition to oscillating mode happens.
\TwoFig{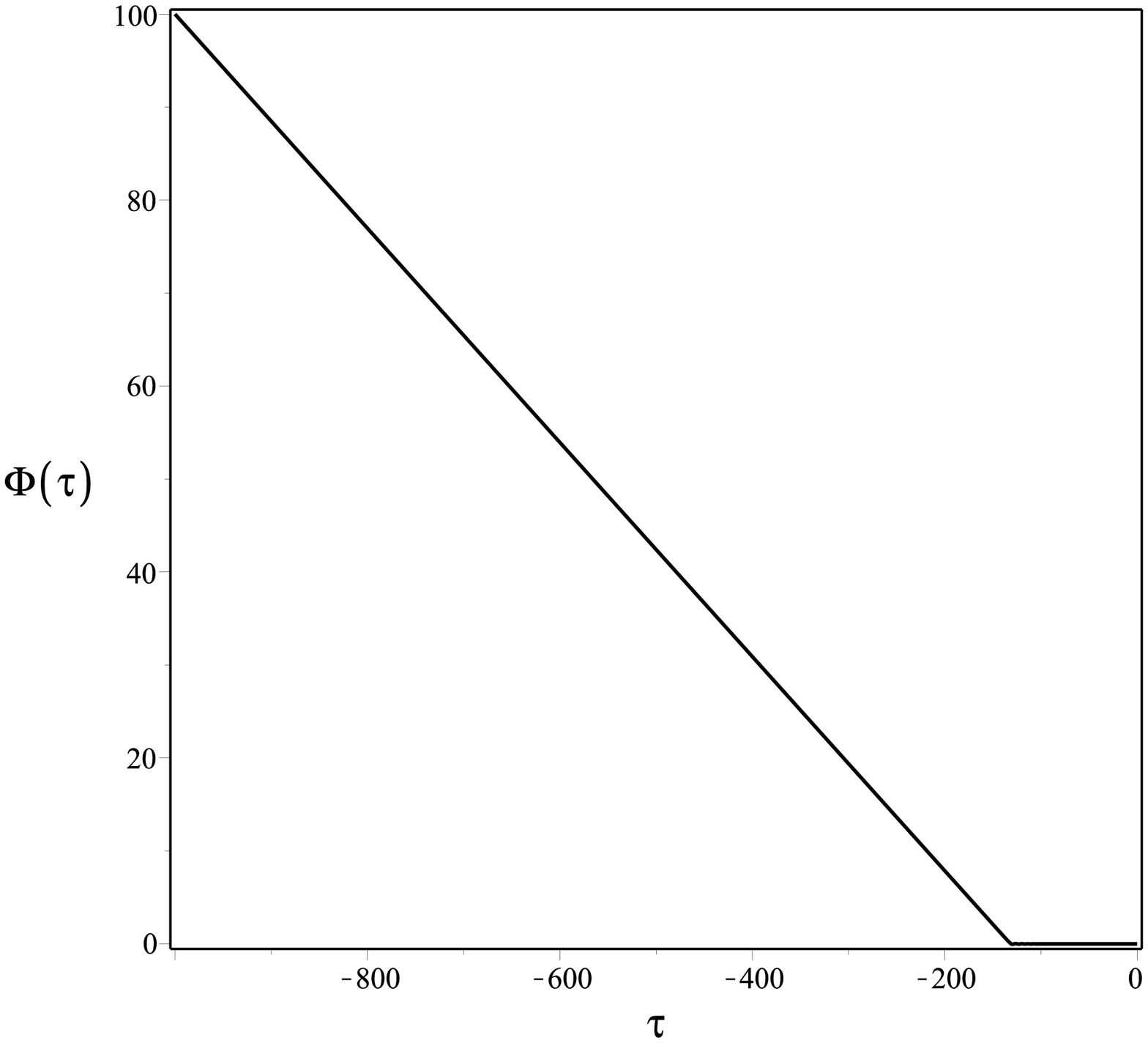}{\label{ris14}The evolution of the potential at early stages at small value of the cosmological constant $\Lambda_m=0.001$;  $\Phi(-1000)=100$.}{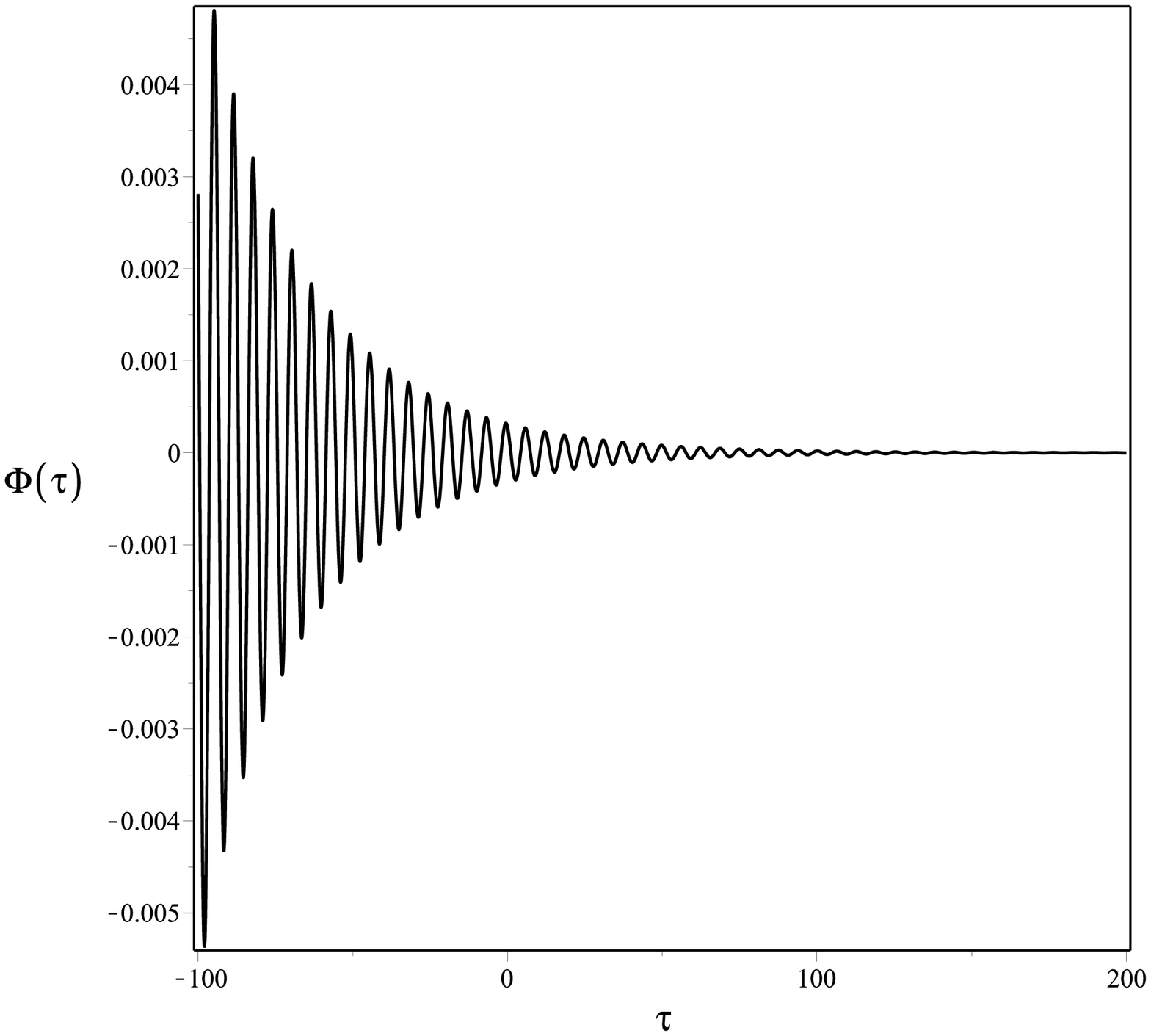}{\label{ris15}The evolution of the potential at late stages at small value of the cosmological constant $\Lambda_m=0.001$;  $\Phi(-1000)=100$.}
Fig. \ref{ris16} -- \ref{ris17} show the evolution of the potential at stage of damped oscillations in the case of zero and small values of the cosmological constant. It is seen that at the same initial values of the scalar potential at this stage the amplitude of the potential's oscil\-la\-tions at the same time is by order of magnitude greater in the case of zero value of the cosmological constant.
\TwoFig{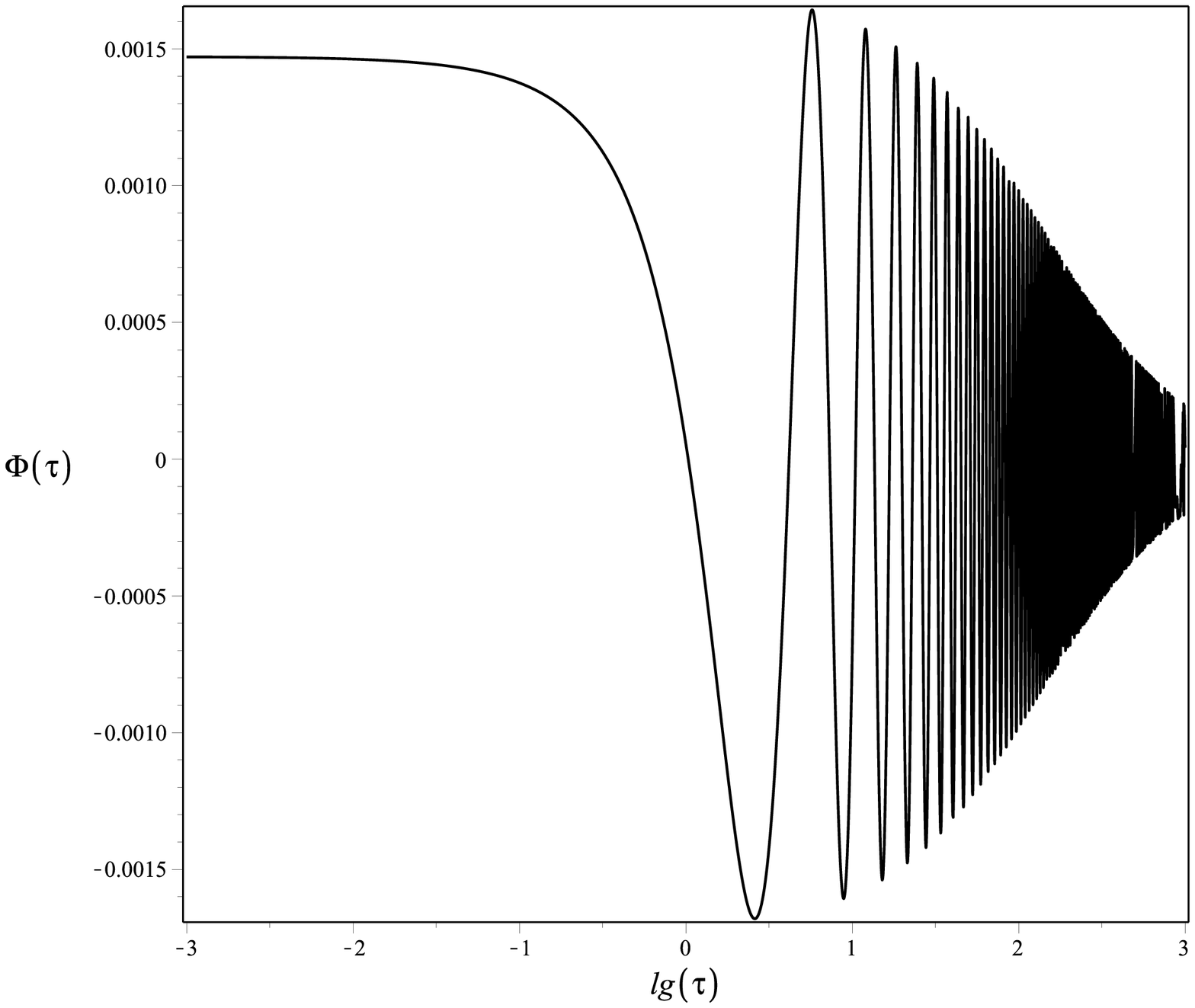}{\label{ris16}The evolution of the potential at stage of oscillations at zero value of the cosmological constant $\Lambda_m=0$ in the logarithmic time scale;  $\Phi(-1000)=100$.}{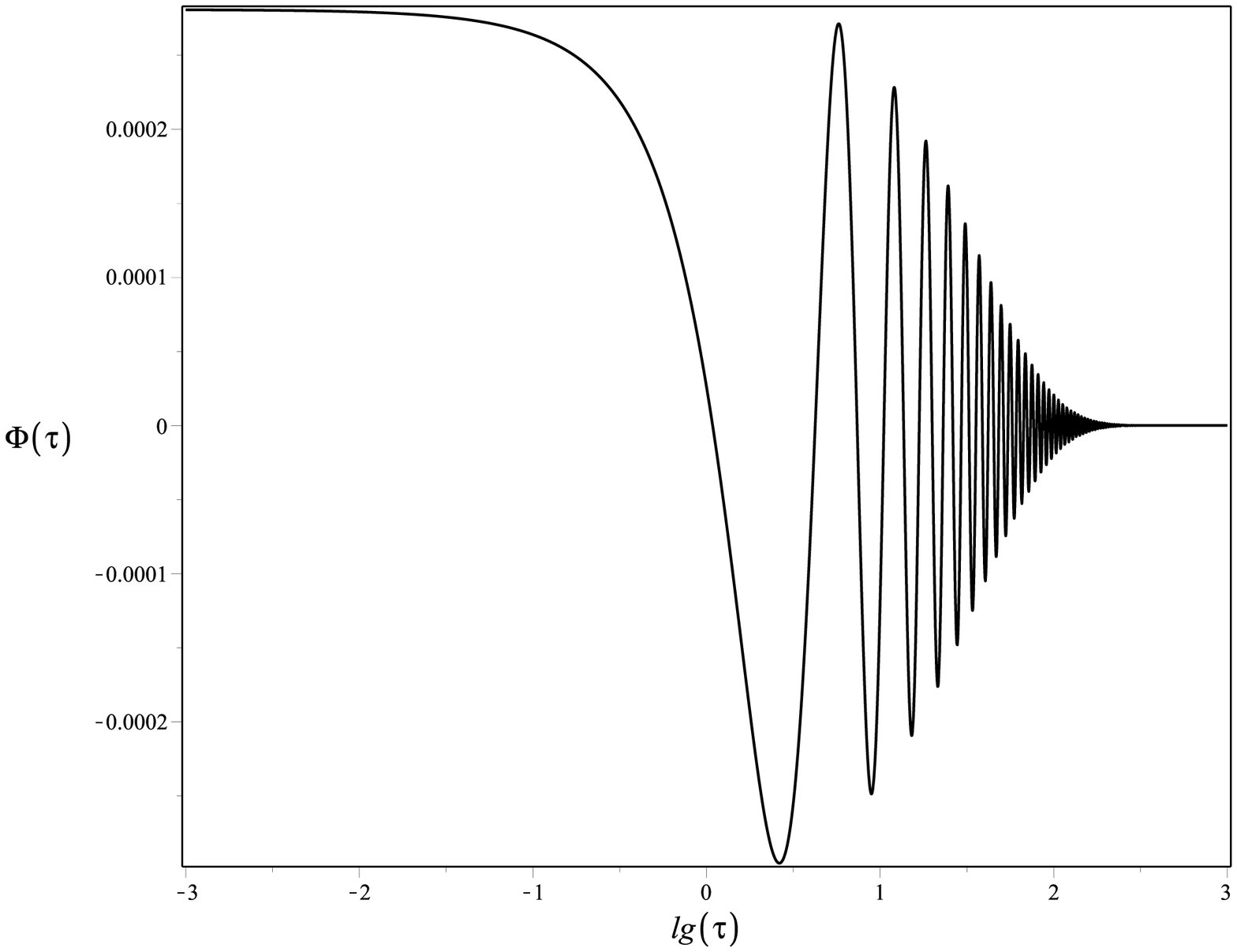}{\label{ris17} The evolution of the potential at stage of oscillations in the logarithmic time scale at small value of the cosmological constant $\Lambda_m=0.001$;  $\Phi(-1000)=100$.}

\subsection{The Evolution of the Hubble Constant $H(t)$}

The fall of the Hubble constant from  $H_1$ (\ref{H_1}) till $H_0$ (\ref{H_0}) starts the later, the bigger is the initial value of the scalar field's potential $\Phi_0=\Phi(-\infty)$ (Fig. \ref{ris18} -- \ref{ris19}).
\TwoFig{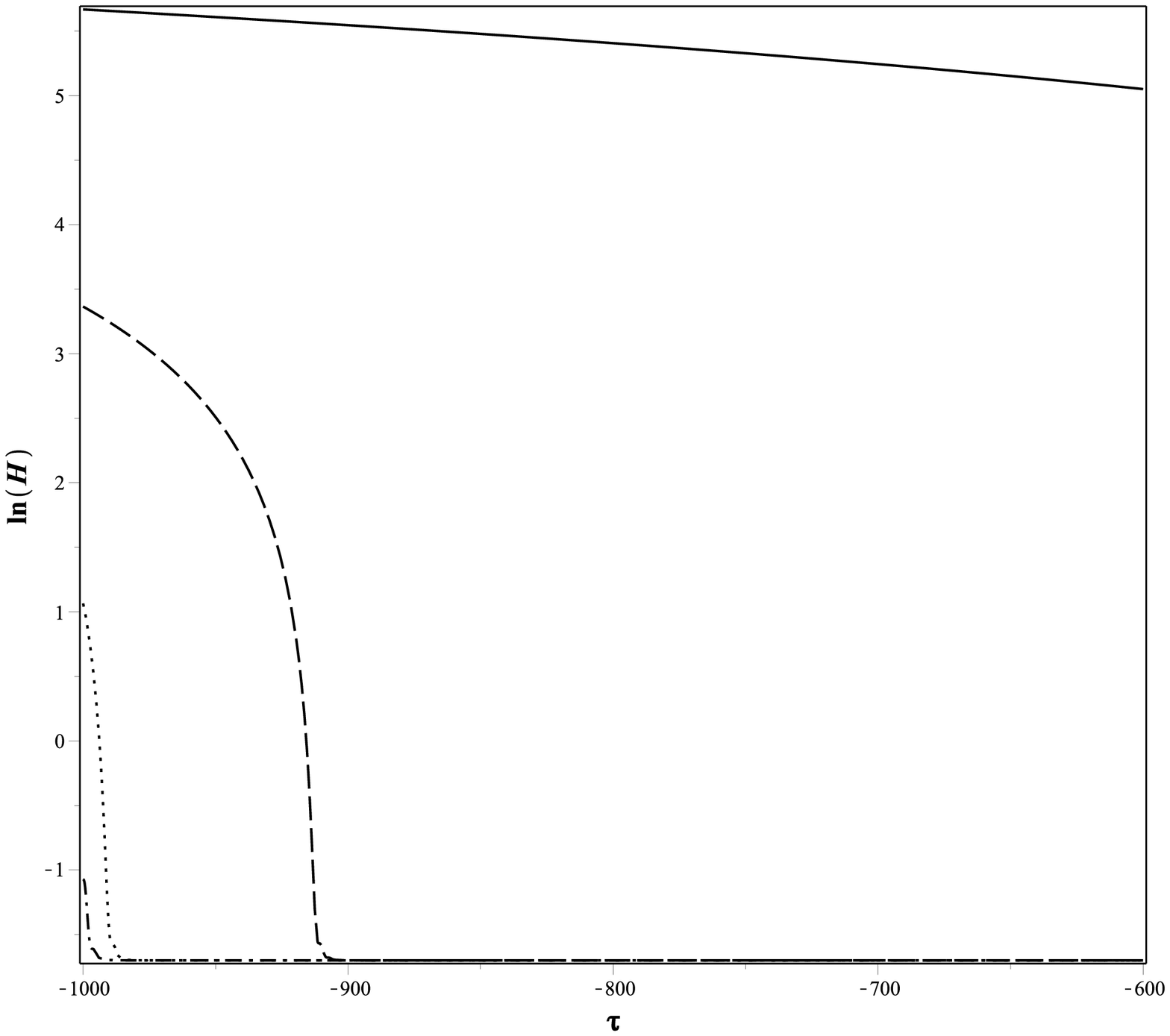}{\label{ris18}The dependency of the Hubble constant's evolution on initial value of the scalar potential: bottom - up:  $\Phi(-1000)=0.1;1;10;30;100$ at small value of the cosmological constant $\Lambda_m=0.01$.}{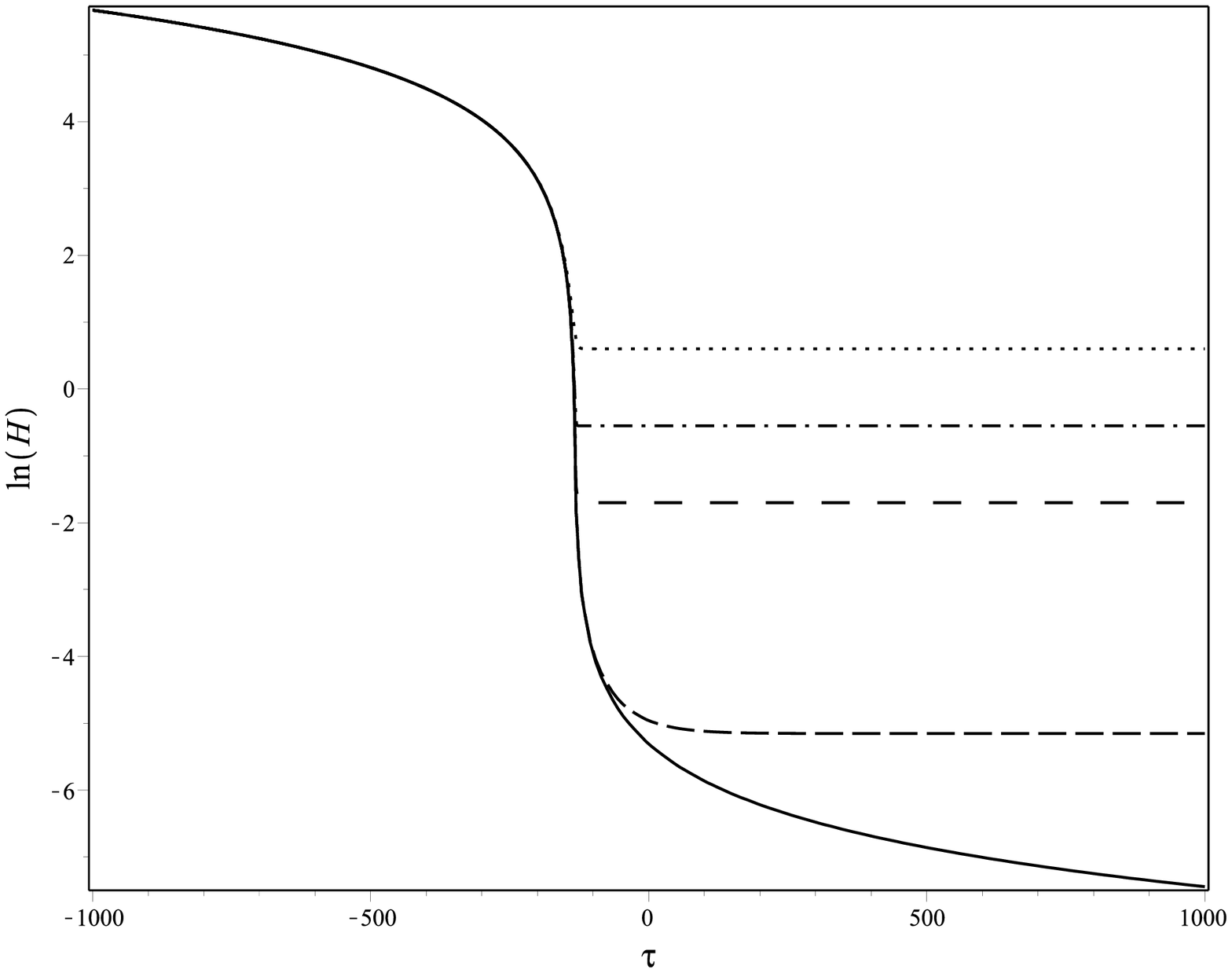}{\label{ris19}The dependency of the Hubble constant's evolution on value of the cosmological constant: bottom - up: $\Lambda_m=0;0.0001;0.1;1;10$ at initial values: $\Phi(-1000)=100$; $Z(-1000)=0$.}
\subsection{The Scale Factor}
The evolution of the scale factor is shown on Fig. \ref{ris20} -- \ref{ris21}. The values of the $\ln$ function are put on the Y-axis of the plots on these pictures
\[L(\tau)=\ln(a(\tau)).\]

Therefore value $\ln L =10$ corresponds to value $L\sim 10^4$ and the value of the scale factor $a/a_0\sim 10^{9566}$.
\TwoFig{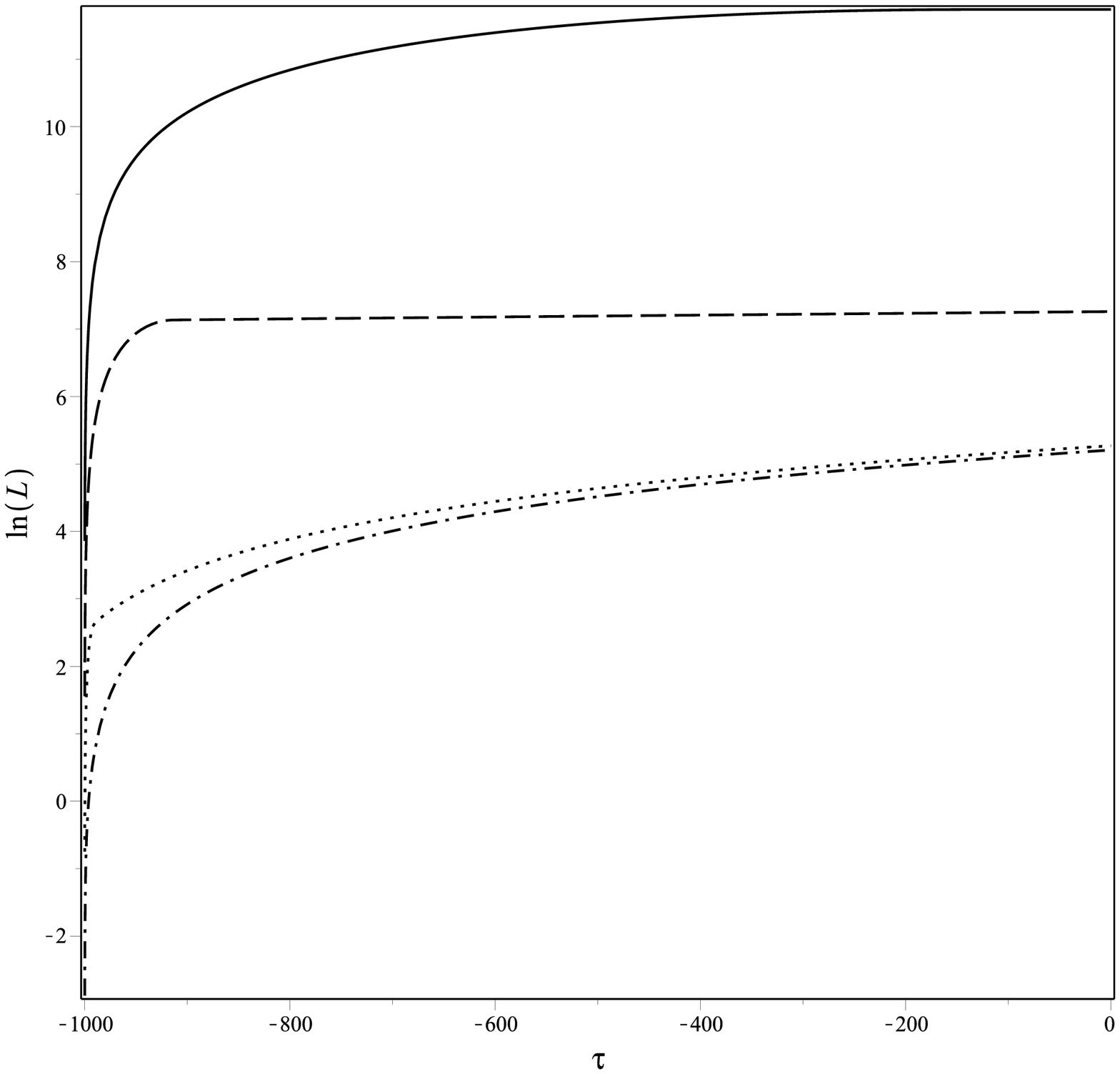}{\label{ris20}The dependency of the scale factor's evolution on the initial value of the scalar potential: bottom-up: $\Phi(-1000)=0.1;1;10;100$ at small value of the cosmological constant $\Lambda_m=0.1$.}{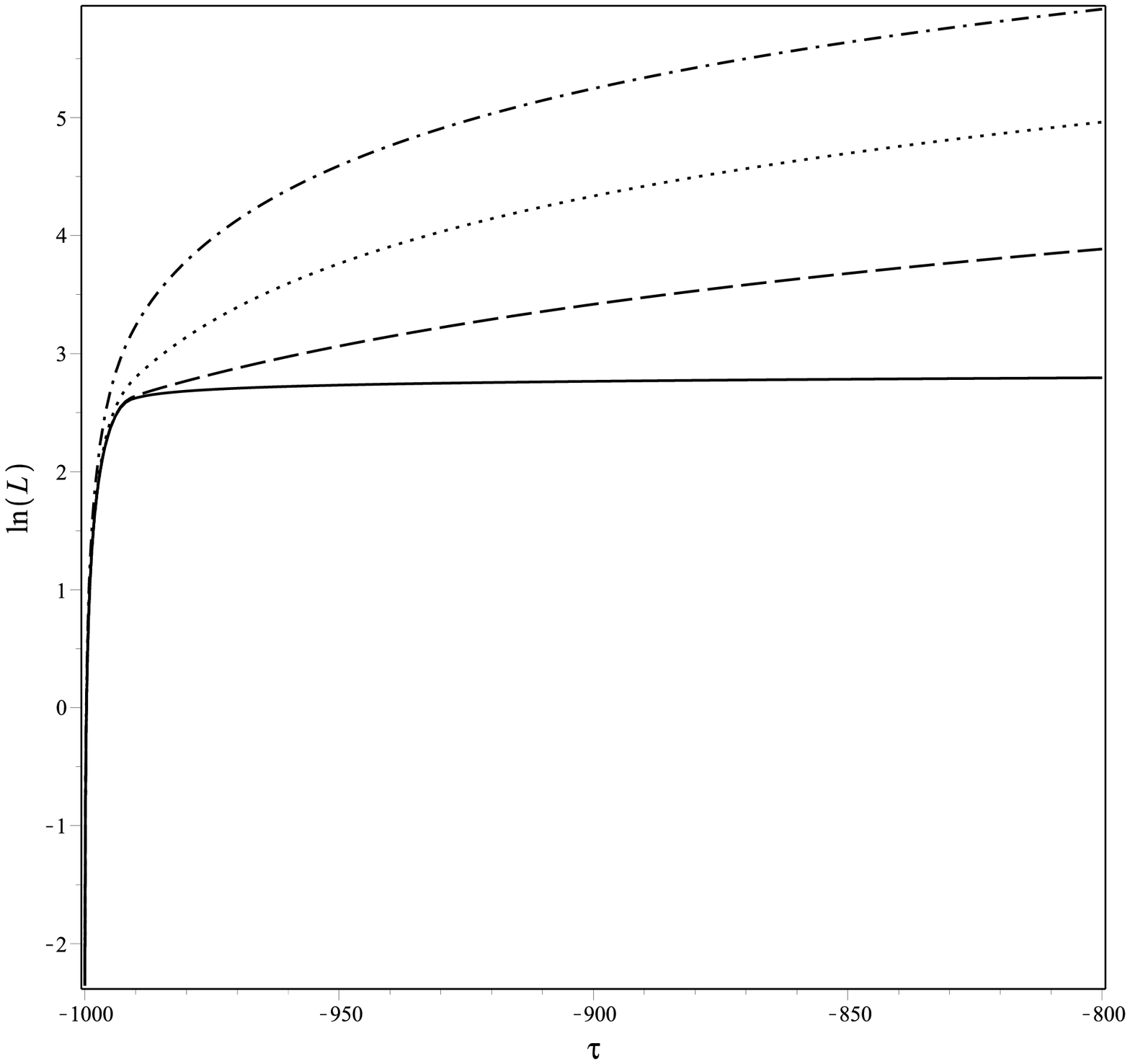}{\label{ris21}The dependency of the scale factor's evolution on the cosmological constant: bottom-up: $\Lambda_m=0;0.1;4/3;10$ at initial values: $\Phi(-1000)=1$; $Z(-1000)=0$.}
\subsection{The Evolution of the Invariant Cosmological Acceleration}
The cosmological acceleration $\Omega$ is calculated using the following formula:
\begin{equation}\label{Omega_h}
\Omega(\tau)=1+\frac{H'_m(\tau)}{H^2_m(\tau)}.
\end{equation}
The value of the cosmological acceleration also oscillates with period $\sim 2\pi$ after the stage of primary inflation. Fig. \ref{ris22} -- \ref{ris23} show the stage of the cosmological acceleration's oscillations. In particular, Fig. \ref{ris23} allows us to see how the average value of the cosmological acceleration grows appro\-xi\-ma\-tely from $-1/2$ to $1$.
\TwoFig{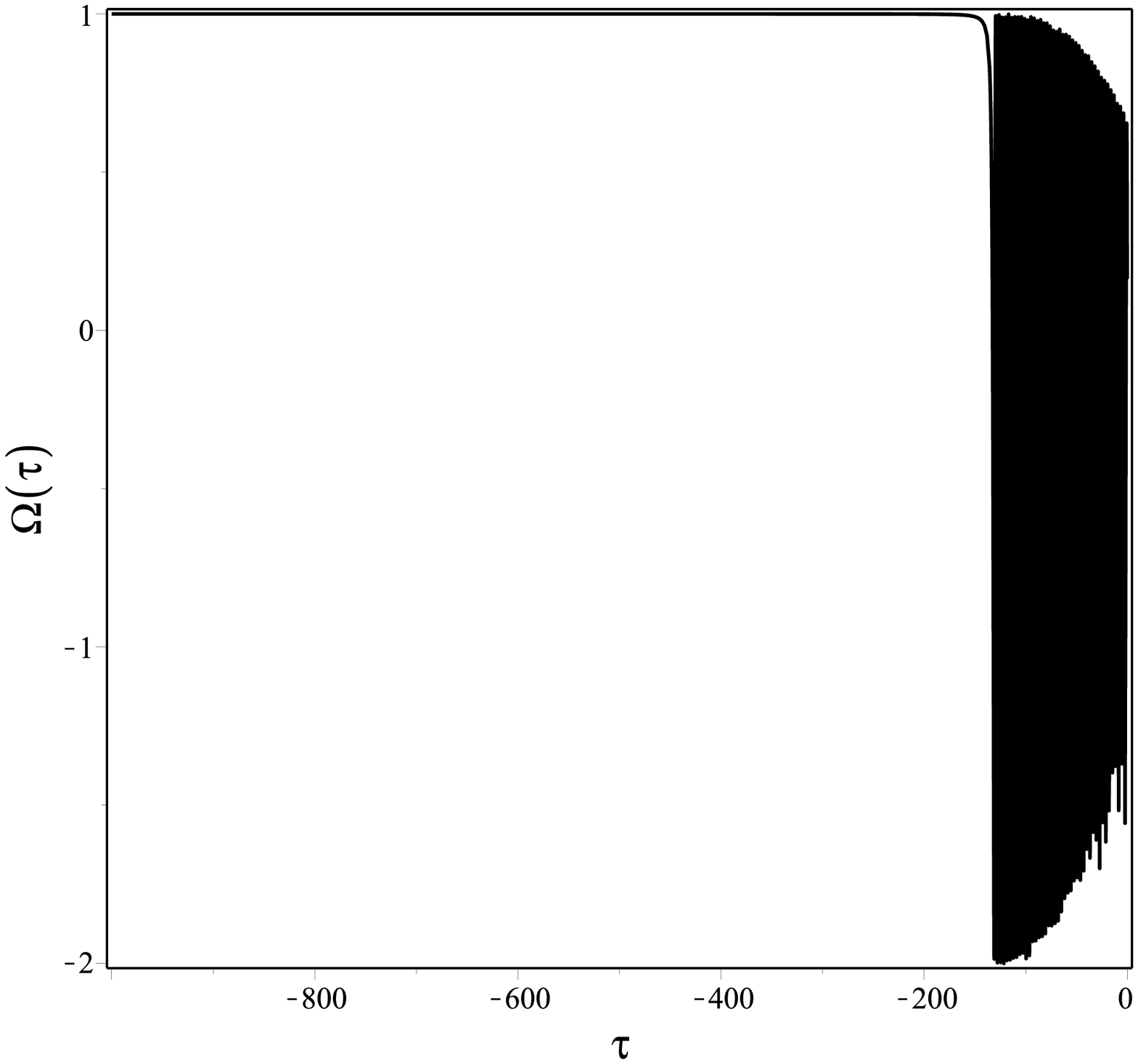}{\label{ris22}The large-scale evolution of the cosmological acceleration: $\Phi(-1000)=100$ at small value of the cosmological constant $\Lambda_m=0.00001$. The black-painted range of the plot represent damped oscillations.}{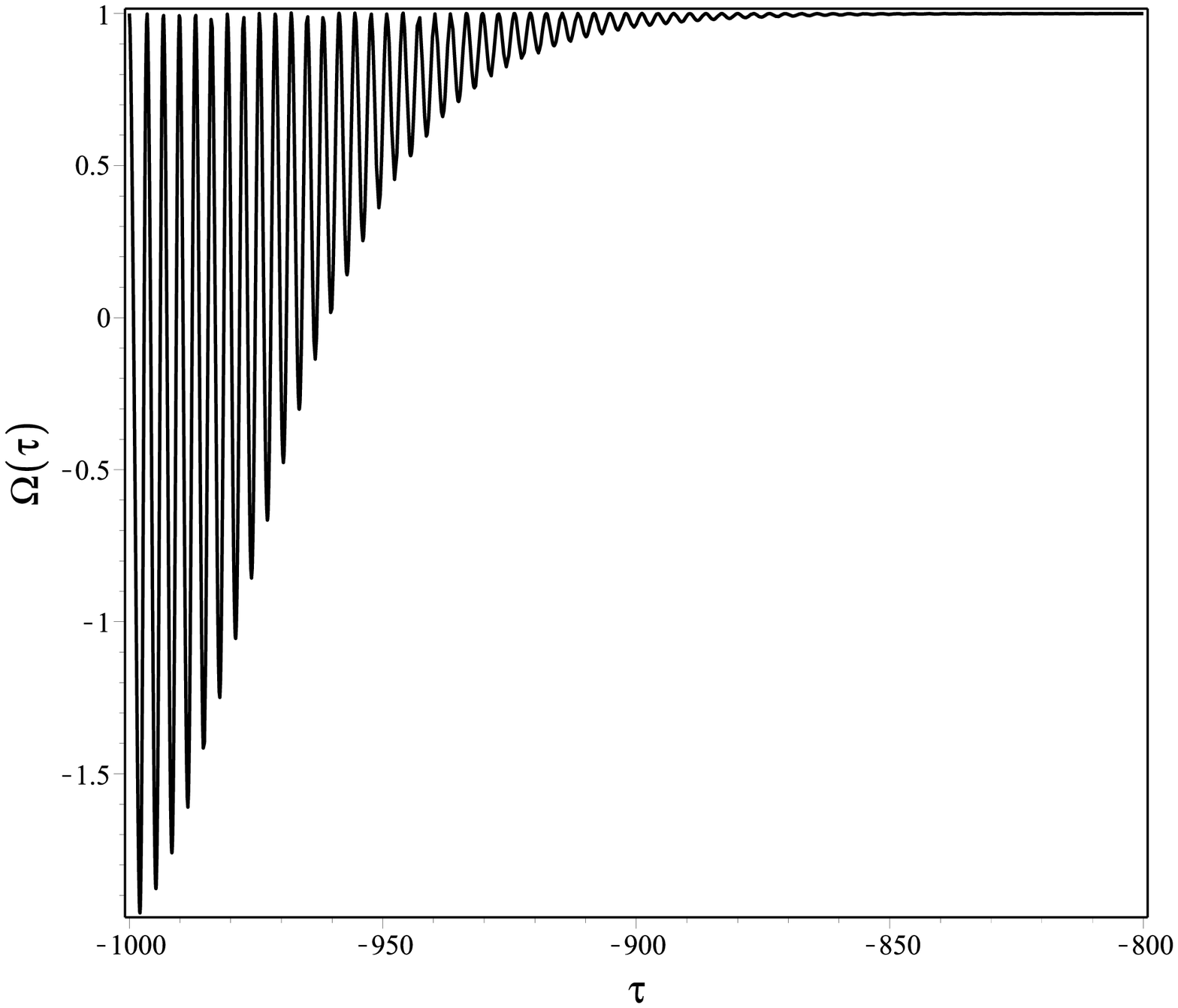}{\label{ris23}The range of oscillating stage of the evolution of the cosmological acceleration $\Lambda_m=0.001$ at initial values: $\Phi(-1000)=0.1$; $Z(-1000)=0$.}
It should be noted that the oscillating character of the cosmological acceleration at late stages for the case of classical massive scalar field also was discovered at numerical integration of the equations of the cosmological model with scalar charged par\-ticles \cite{Yu_AAS}.

\subsection{The Average of the Cosmological Acceleration}

The fact that the invariant cosmological acceleration has an oscillating character at great times (Fig. \ref{ris25}), having the period of these oscillation on the time scale $\tau$ of the order of $2\pi$, i.e. in the ordinary time scale $T\sim2\pi/m$ being an obviously microscopic value, leads to necessity of introduction of the average value of the invariant cosmological acce\-le\-ra\-tion averaged by large enough number of oscillations i.e. by large enough interval $\Delta\tau=N\cdot2\pi$, where $N \gg1$:

\begin{equation}\label{Omega_average0}
\overline{\Omega(\tau,\Delta\tau)}\equiv \frac{1}{\Delta\tau}\int\limits_{\tau}^{\tau+\Delta\tau}\Omega(\tau')d\tau'.
\end{equation}
Using (\ref{Omega_average0}) in formula (\ref{Omega_h}) and carrying out ele\-men\-tary integration, let us find  the following expression for the average cosmological acceleration:
\begin{equation}\label{Omega_average}
\overline{\Omega(\tau,\Delta\tau)}=1+\frac{1}{\Delta\tau}\biggl(\frac{1}{h(\tau)}-\frac{1}{h(\tau+\Delta\tau)}\biggr).
\end{equation}
\TwoFig{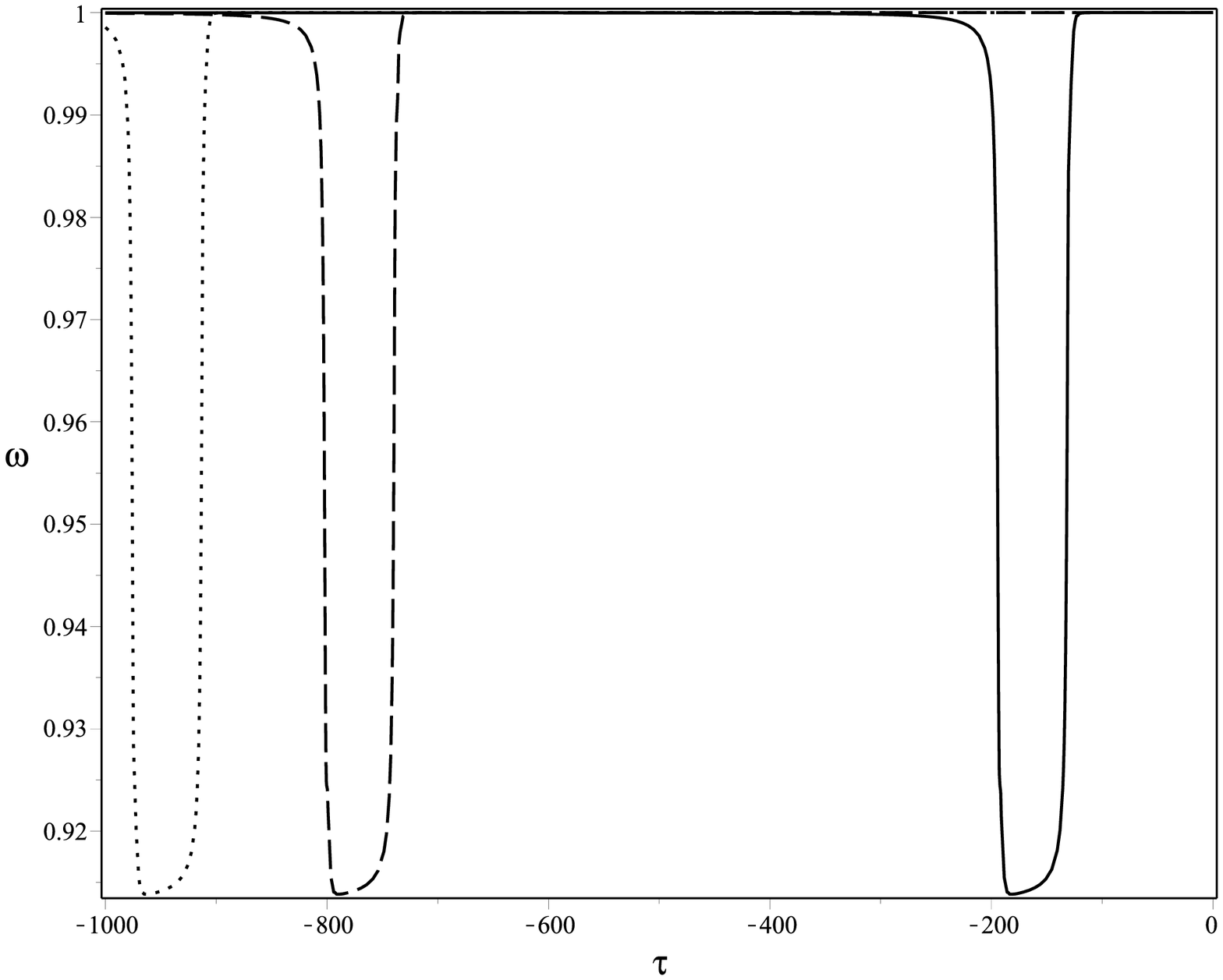}{\label{ris24}The evolution of the average cosmological acceleration: $\Phi(-1000)=100$ at cosmological constant $\Lambda_m=0.1$. The duration of the interval of averaging $\Delta\tau=20\pi$.}
{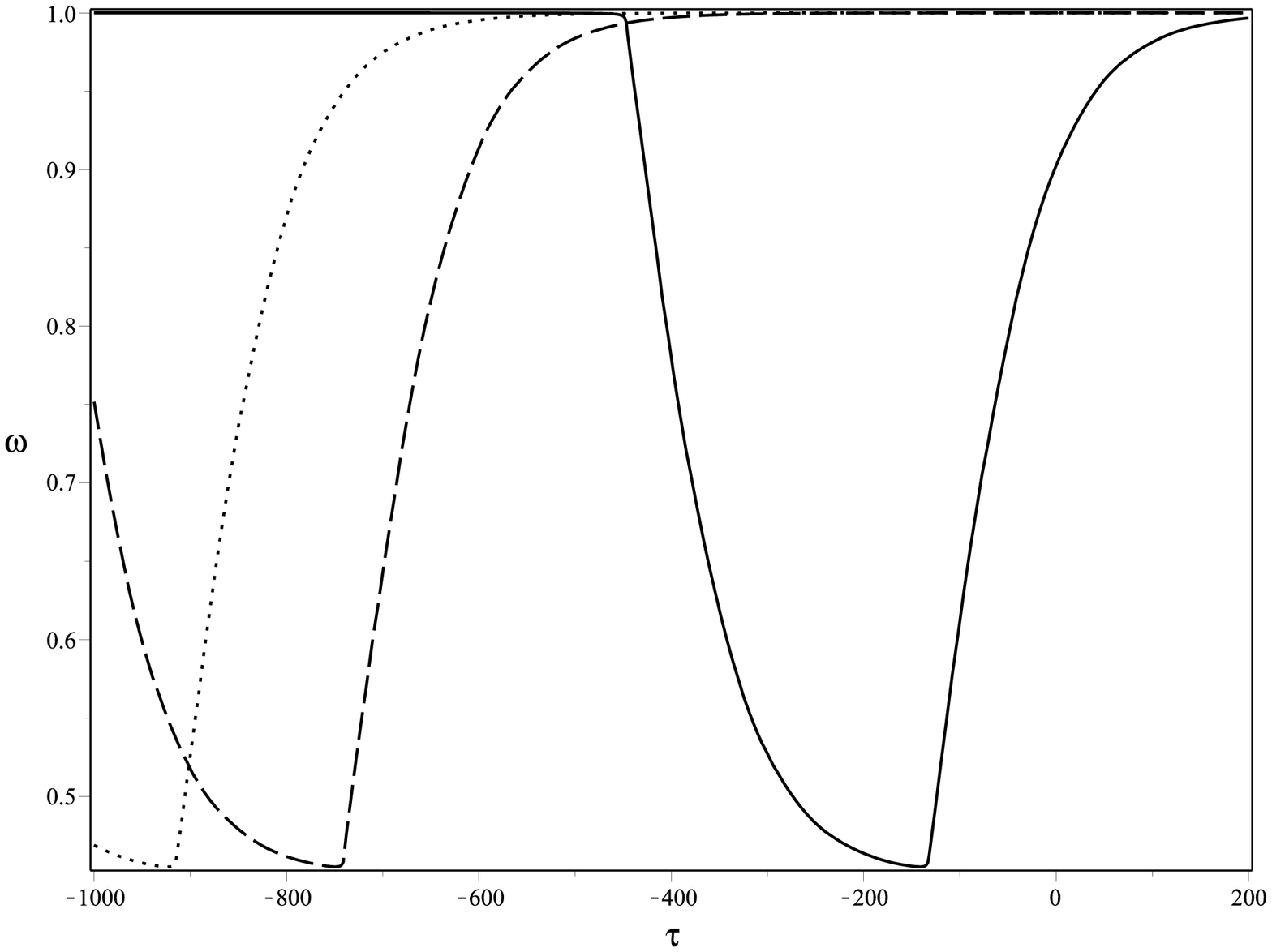}{\label{ris25}The evolution of the average cosmological acceleration: $\Phi(-1000)=100$ at cosmological constant $\Lambda_m=0.0001$. The duration of the interval of averaging $\Delta\tau=100\pi$.}
\TwoFig{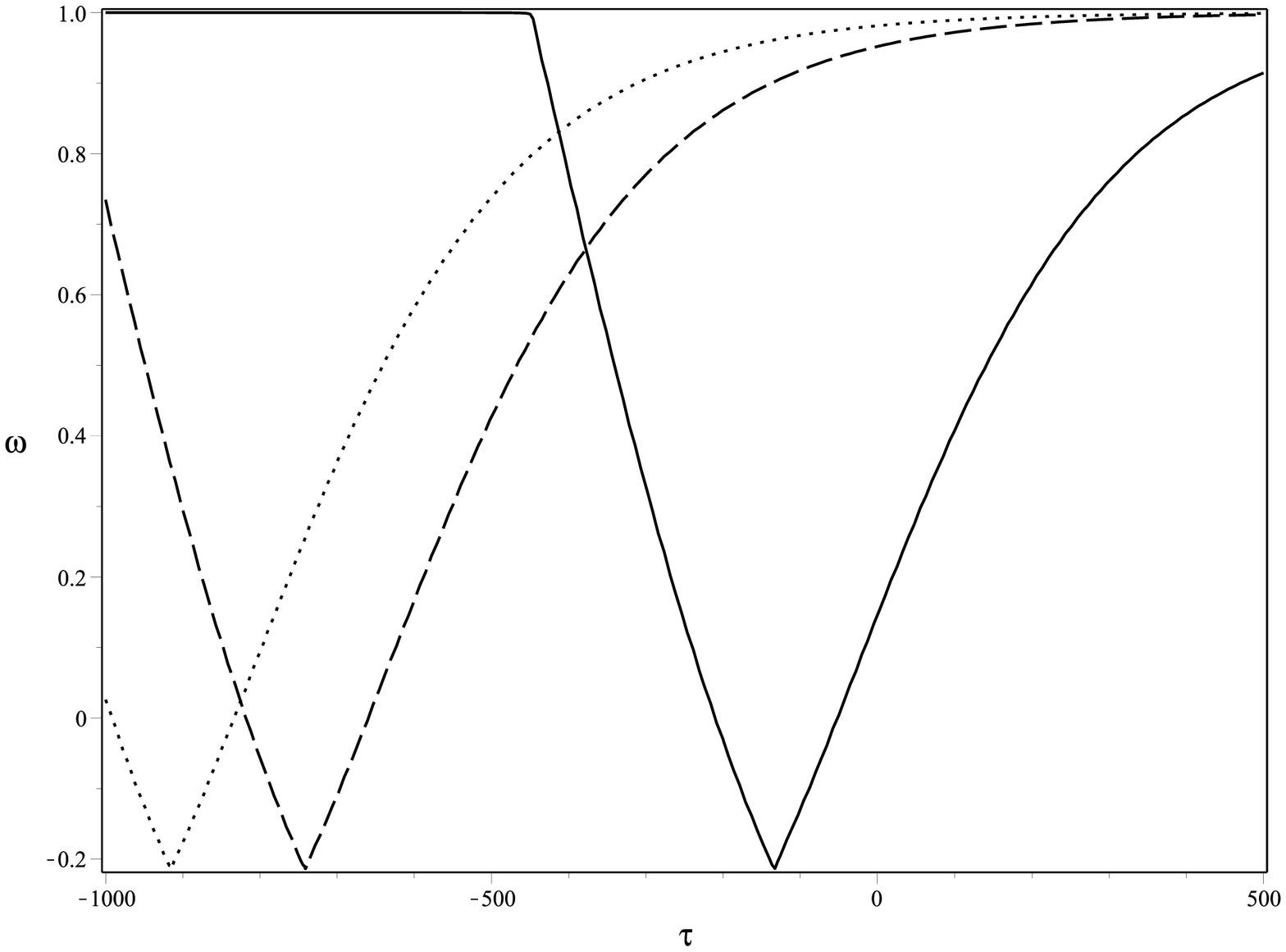}{\label{ris26}The evolution of the average cosmo\-logical acceleration: $\Phi(-1000)=100$ at cosmological constant $\Lambda_m=0.00001$. The duration of the interval of averaging $\Delta\tau=100\pi$.}
{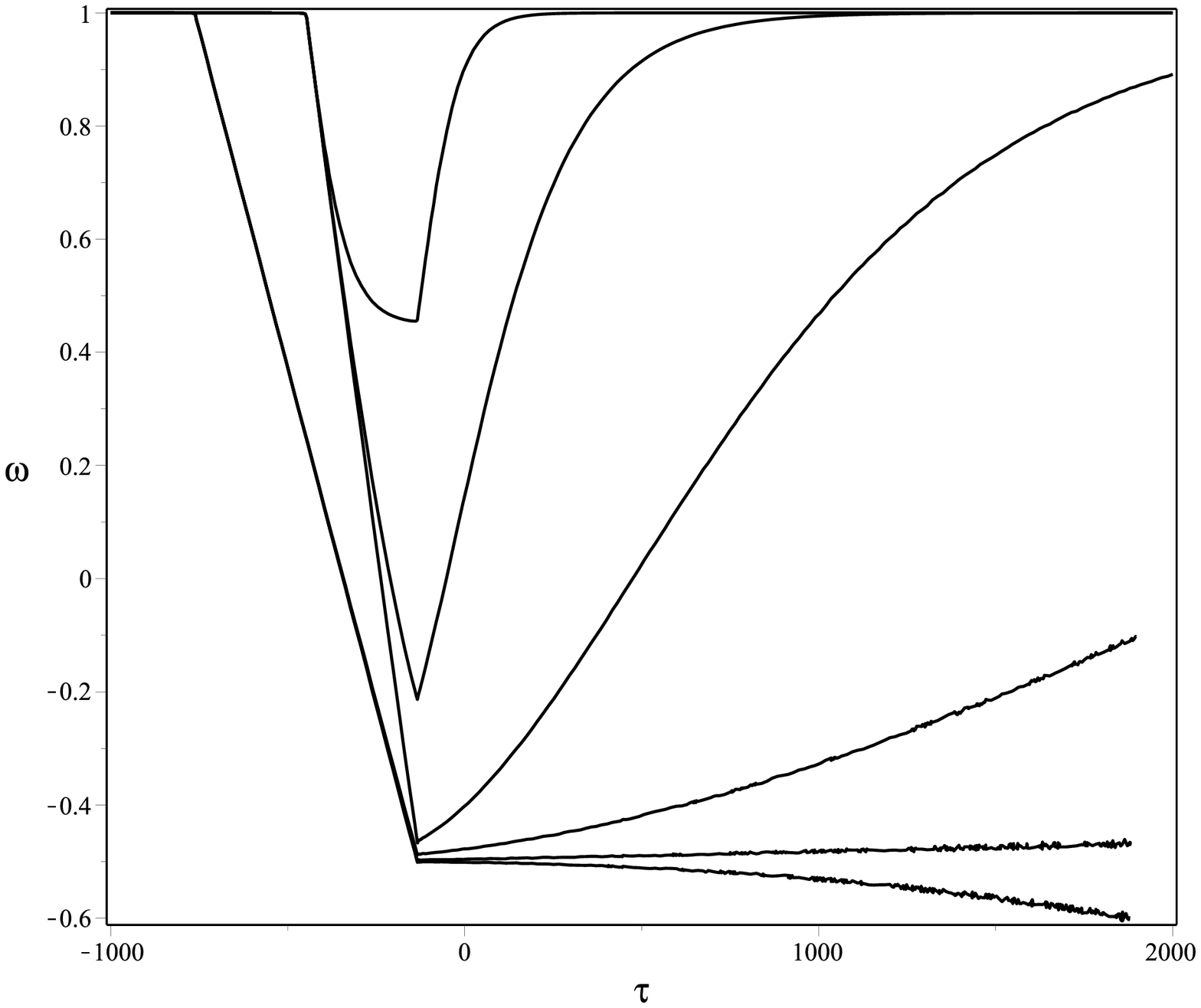}{\label{ris27}The evolution of the average cosmo\-lo\-gical acceleration: $\Phi(-1000)=100$ at cosmological constant -- bottom-up: $\Lambda_m=10^{-8};3\cdot 10^{-8}; 10^{-7};
10^{-6}; 10^{-5}; 10^{-4}$. The duration of the interval of averaging $\Delta\tau=100\pi\div 200\pi$.}
\section{The Conclusion}
Thus, we can state a fact that at small values of the cosmological constant $\Lambda_m\leq3\cdot10^{-8}$ the evolution of the dynamic system laid in the basis of SCM, at large time intervals very weekly differ from the evolution of the dynamic system without account of the cosmological constant. In particular, sufficiently long non - relativistic stage appears  in that system.

It must be noted that if using the standard model of the elementary particles with mass of Higgs boson of order of 10 Tev $\sim 10^{-15}m_{pl}$ in the capacity of the base model , this limitation gives us $\Lambda\leq 3\cdot10^{-38}$. As is known, the value of the cosmological constant is estimated as $10^{-123}$, so that the real cosmological situation relates exactly to the considered case, which corresponds to the first kind of a the singular point -- attractive pole. As we noted above, the final stage of such a cosmo\-lo\-gical model is the inflation one while at inter\-me\-diate stages of expansion there automatically appears the non-relativistic mode.

In conclusion, the Authors express their gratitude to the members of MW seminar for relativistic kinetics and cosmology of Kazan Federal University for helpful discussion of the work.

%%%%%%%%%%%%%%%%%%%%%%%%%%%%%%%%%%%%%%%%%%%%%%%%%%%%%%%%%%%%%%%%%
%%%%%%%%%%%%%%%%%%%%%%%%%%%%%%%%%%%%%%%%%%%%%%%%%%%%%%%


\begin{thebibliography}{4}
%
\bibitem{Yu_Quality}
Yurii Ignat'ev,	arXiv:1609.00745 [gr-qc].
%
\bibitem{Gorb_Rubak}
D.S. Gorbunov and V.A. Rubakov, Introduction to the Theory of the Early Universe: Cosmological
Perturbations and Inflationary Theory. Singapore: World Scientific (2011).
%
\bibitem{Bogoyavlensky}
O.I. Bogoyavlensky, Methods of the qualitative theory of dynamical systems in astrophysics and gas dynamics. Moskow, Nauka, (1980).
%
\bibitem{Yu_AAS}
Yurii Ignat'ev, Alexander Agathonov, Mikhail Mikhailov and Dmitry Ignatyev, Astrophys Space Sci (2015) 357:61; arXiv:1411.6244v1 [gr-qc].
%
\end{thebibliography}
\end{document}